%% file: paper_draft.tex
\documentclass[a4paper,useAMS,usenatbib]{mnras}
\usepackage{graphics,epsfig,psfig}
\hypersetup{colorlinks=true,allcolors=blue}
\usepackage{todonotes}
\usepackage[normalem]{ulem}
\usepackage{xcolor}
\usepackage{float}
\usepackage{booktabs}
\usepackage[style=base, singlelinecheck=off, font=small, labelfont=bf, 
justification=raggedright]{caption}
\usepackage{subfigure}
\usepackage{orcidlink}
\usepackage{multicol}
\usepackage{longtable}
\usepackage{supertabular}
\usepackage{graphicx}
\usepackage[]{inputenc,amssymb}
\usepackage{natbib}
\usepackage{url}
\usepackage{widetext}
\usepackage{amsmath}
\usepackage{cancel}

  \makeatletter
\def\@fnsymbol#1{\ensuremath{\ifcase#1\or \dagger\or \ddagger\or
   \mathsection\or \mathparagraph\or \|\or **\or \dagger\dagger
   \or \ddagger\ddagger \else\@ctrerr\fi}}
    \makeatother
\def \be{\begin{equation}}
\def \ee{\end{equation}}
\def \bea{\begin{eqnarray}}
\def \eea{\end{eqnarray}}

\def\mnras{MNRAS}

\definecolor{webgreen}{rgb}{0,.5,0}
\definecolor{webbrown}{rgb}{.6,0,0}

\title[BBHs population and cosmology]{Binary black holes  population and cosmology in new lights: Signature of PISN mass and formation channel in GWTC-3}

\author[Karathanasis, Mukherjee, Mastrogiovanni]{Christos Karathanasis$^{1}$\thanks{ckarathanasis@ifae.es}\orcidlink{0000-0002-0642-5507}, Suvodip Mukherjee$^{2,3}$\thanks{suvodip@tifr.res.in}\orcidlink{0000-0002-3373-5236}, Simone Mastrogiovanni$^{4}$\thanks{simone.mastrogiovanni@oca.eu}\orcidlink{0000-0003-1606-4183}
\\
$^{1}$ Institut de Física d’Altes Energies (IFAE), Barcelona Institute of Science and Technology, Barcelona, Spain\\ 
$^{2}$ Perimeter Institute for Theoretical Physics, 31 Caroline Street N., Waterloo, Ontario, N2L 2Y5, Canada\\
$^{3}$ Department of Astronomy \& Astrophysics, Tata Institute of Fundamental Research, 1, Homi Bhabha Road, Colaba, Mumbai 400005, India\\
$^{4}$ Artemis, Université Côte d’Azur, Observatoire de la Côte d’Azur, CNRS, F-06304 Nice,
France\\}

\pubyear{2022}
\begin{document}

\pagerange{\pageref{firstpage}--\pageref{lastpage}}
\maketitle

\label{firstpage}

\begin{abstract}
The mass, spin, and merger rate distribution of the binary black holes (BBHs) across cosmic redshifts provide a unique way to shed light on their formation channel. Along with the redshift dependence of the BBH merger rate, the mass distribution of BBHs can also exhibit redshift dependence due to different formation channels and dependence on the metallicity of the parent stars. We explore the redshift dependence of the BBH mass distribution jointly with the merger rate evolution from the third gravitational wave (GW) catalog GWTC-3 of the LIGO-Virgo-KAGRA collaboration.  We study possible connections between peak-like features in the mass spectrum of BBHs and processes related to supernovae physics and time-delay distributions. 
 We obtain a preference for short-time delays between star formation and BBH mergers. Using a power law form for the time delay distribution ($(t^{\rm min}_d)^{d}$) we find $d<-0.7$ credible 
 at 90\% interval.
The mass distribution of the BBHs  {could be fitted} with a power-law form with  {a redshift-dependent peak feature that can be linked  to the pair instability supernovae (PISN) mass scale $M_{\rm PISN}(Z_*)$ at a stellar metallicity $Z_*$. For a fiducial value of the stellar metallicity $Z_*= 10^{-4}$, we find the $\rm M_{\rm PISN}(Z_*)=44.4^{+7.9}_{-6.3} $ $\rm M_\odot$.} This is in accordance with the theoretical prediction of the lower edge of the PISN mass scale and differs from previous analyses. 
 Although we find a strong dependence of the PISN value on metallicity, the model that we explored is not strongly favored over those that do not account for metallicity as the Bayes factors are inconclusive. In the future with more data, evidence towards metallicity dependence of the PISN will have a significant impact on our understanding of stellar physics.
\end{abstract}

\begin{keywords} 
gravitational waves, black hole mergers, cosmology: miscellaneous
\end{keywords}
\section{Introduction}
Gravitational wave (GW) observations bring a wealth of information to a broad range of topics ranging from astrophysics, cosmology, and fundamental physics. The first GW detection\citep{Abbott:2016:gw150914} opened a new way of observing the Universe. The latest measurements from the LIGO-Virgo-KAGRA (LVK)\citep{Harry:2010,Aso:2013,Acernese:2014,LIGOScientific:2014pky,Martynov:2016fzi,KAGRA:2013pob,Tse:2019,KAGRA:2020tym,gwtc2.1} have detected $90$ compact objects which constituents of binary neutron stars (BNSs), neutron star binary black holes (NSBHs), and binary black holes (BBHs) \citep{LIGOScientific:2021psn}. The observed GW sources give a direct probe to infer the mass distribution of the compact objects across a range of cosmic redshifts. The recent measurement by the LVK collaboration exhibit that the mass distribution of the BBHs shows a power-law+Gaussian (PLG) distribution \citep{Abbott:2019:population,Abbott_2021, LIGOScientific:2021psn,LIGOScientific:2021djp,Virgo:2021bbr}. Along with the mass distribution, a power-law model of the BBHs merger rate is inferred from LVK analysis \citep{Abbott_2019,Abbott_2021, LIGOScientific:2021psn,LIGOScientific:2021djp,Abbott2020b}. The mass distribution of GW sources and the merger rate provides a direct way to understand the formation channel of BBHs if an underlying physical model can be inferred from observations.  

The currently used phenomenological PLG model of mass distribution does not consider redshift evolution. However, the mass distribution of the astrophysical BBHs is likely to exhibit a redshift dependence due to the dependence of the black hole masses on stellar properties, such as the stellar metallicity \citep{Bethe:1998bn, 1998A&A...332..173P,2002ApJ...567..532H,2002ApJ...572..407B,2012ApJ...759...52D, Dominik:2014yma, Mapelli:2017hqk,Spera:2017fyx,2018MNRAS.474.2959G,Toffano:2019ekp,2019ApJ...887...53F, Renzo:2020rzx,Baxter:2021swn, 2022ApJ...924...39M}. One of the inevitable ways the BBHs distribution can get a  {complex} redshift  dependence  {is through}  {a time delay acting between the binary formation and the merger.} \citep{Mukherjee:2021rtw}.  {This delay causes the population of BBHs at a given redshift to encode information of astrophysical processes and channels that were present at a different cosmic epoch}.  {The time delay contribution will produce a BBH population at a merger redshift $z$ (hat is non-trivial to describe with simple models)} which is composed by black holes formed  {at different cosmic times} with possibly different astrophysical formation channels.  {The time delay distribution and the dependence of the astrophysical processes on cosmic time can lead to a non-trivial BBH merger mass spectrum.} 

The mass distribution of the BBHs can also play an important role in inferring the cosmic expansion history  \citep{Taylor:2011fs,Farr:2019twy, Mastrogiovanni:2021wsd,You:2020wju,Mancarella:2021ecn,Mukherjee:2021rtw,Leyde:2022orh,Ezquiaga:2022zkx}. As the masses of the GW sources are redshifted ($m^{\rm det}= (1+z)m$), one can expect to infer the redshift from the mass distribution of the GW sources, if the mass distribution of the BBHs can exhibit a universal property or at least a standardized behavior. We can break the mass-redshift degeneracy and infer the cosmic expansion history from dark standard sirens without applying the cross-correlation technique  \citep{PhysRevD.93.083511,Mukherjee:2018ebj, Mukherjee:2019wcg, Mukherjee:2020hyn,Bera:2020jhx, Scelfo:2020jyw, Mukherjee:2020mha,2021ApJ...918...20C,Scelfo:2021fqe,2022MNRAS.511.2782C,Mukherjee:2022afz} or statistical host identification technique \citep{Schutz,MacLeod:2007jd,DelPozzo:2012zz,Arabsalmani:2013bj,  PhysRevD.101.122001, Fishbach:2018gjp, Abbott:2021:H0, Soares-Santos:2019irc, Finke:2021aom, Abbott:2020khf, Palmese:2021mjm}. However, if  BBHs mass distribution exhibits redshift dependence due to its intrinsic dependence on the delay time distribution, then cosmic redshifts cannot be accurately inferred \citep{Mukherjee:2021rtw,Ezquiaga:2022zkx}, and it can bias the results if the redshift dependence of the mass distribution is not considered. Exploring the merger rate distribution to explore cosmology from dark sirens is also studied for the third generation GW detectors \citep{Ding:2018zrk, Ye:2021klk, Leandro:2021qlc}. 

In this paper, we make a first joint estimation of the BBH merger rate evolution, mass distribution,
 {and metallicity dependance parameters, by} allowing for the redshift dependence of the BBH mass distribution and $H_0$. This measurement makes it possible to also infer the delay time distribution of the BBHs in a consistent framework along with the cosmological parameters. 
The paper is organized as follows, In Sec. \ref{mass-dist} we discuss the redshift dependence of the mass distribution of the BBHs and its merger rate. In Sec. \ref{bayes} and Sec. \ref{results} we discuss the basic Bayesian framework used in this analysis, the results from the joint estimation and we compare these results with the results inferred by the LVK collaboration. Finally, we conclude the analysis of this work and future prospects in Sec. \ref{conc}.  

\section{Modelling redshift dependence of the binary black hole source population}\label{pop-dist}
In this analysis we use the following model to describe the distribution of BBHs in terms of their source frame masses $m_1$, $m_2$, and merger redshift $z_m$:
\begin{equation}\label{eq:pop_model_}
p(m_1,m_2,z_m|\Phi)=p(m_1,m_2|z_m,\Phi_{m},\Phi_{d},\Phi_{\rm nuis})p(z_m|\Phi_d,\Phi_c).
\end{equation}
where $\Phi=\{\Phi_{\rm m}, \Phi_{\rm c}, \Phi_{\rm d},\Phi_{\rm nuis}\}$ are  a set of population parameters governing the mass model ($\Phi_{\rm m}$), cosmology ($\Phi_{\rm c}$), time delay ($\Phi_{\rm d}$) and a set of nuisance parameters ($\Phi_{\rm nius}$). In the following section we explain in detail each of those terms individually.

\subsection{The redshift dependence of the mass distribution: Mixing of black holes}\label{mass-dist}
The distribution of BBHs observed by LVK spans a range of redshift and masses which is currently modeled using different phenomenological models out of which the PLG model fits the data the best \citep{Abbott:2019:population,Abbott_2021, LIGOScientific:2021psn,LIGOScientific:2021djp,Virgo:2021bbr} but does not explore the redshift dependence of the BBH mass distribution. In this paper, we explore how an astrophysically motivated mass distribution of the BBHs originating due to the effect of the \textit{time delay distribution} agrees with the GWTC-3 data.
The observed mass distribution of the BBHs is driven by the underlying astrophysical properties of the parent stars of the individual black hole and the mixing of black holes formed in different redshifts, both of which lead to a redshift dependence of the observed BBH mass distribution.

The redshift dependence of the observed BBH mass distribution in the mixing of binary black hole model, that we consider in this analysis, is due to three effects, (i) metallicity dependence of the pair-instability supernovae (PISN) mass scale, (ii) redshift evolution of the stellar metallicity, (iii) distribution of delay times between the formation of the stars that will later become BHs, and for them to merge with another black hole. We will briefly describe below all these aspects. 

(i) According to the PISN process, the mass distribution of BHs is expected to feature a mass gap due to the mass loss of heavy stars \citep{Spera:2017fyx,2019ApJ...887...53F, Renzo:2020rzx}. The mass loss during the PISN sets the lower limit of the mass gap at around \textit{$M_{\rm PISN}=45$} M$_\odot$. However, $M_{\rm PISN}$ is also closely related to the stellar metallicity. It was shown \citep{Spera:2017fyx,2019ApJ...887...53F, Renzo:2020rzx} that $M_{\rm PISN}$ varies less than 10$\%$ for a variation of the stellar metallicity Z from $10^{-5}$ to $3\times 10^{-3}$ from a 1-D stellar evolution model Modules for Experiments in Stellar Astrophysics (MESA) \citep{2011ApJS..192....3P,2019ApJS..243...10P}. The stars with higher metallicity have a larger mass loss due to stellar winds, which leads to a lower value of the PISN mass scale, than stars formed with lower metallicity. As a result, the position of the $M_{\rm PISN}$ will vary.  {Given the simplicity of current modeling of stellar winds in 1-D codes such as MESA and the lack of independent observations to determine the PISN mass scale, the dependence of $M_{\rm PISN}$ on stellar properties is still subject to large uncertainties.} 

(ii) The metallicity in the Universe varies with redshift and also with the individual galaxies. The global evolution of the stellar metallicity \citep{2002ApJ...572..407B,2012ApJ...759...52D, Dominik:2014yma, Mapelli:2017hqk,2018MNRAS.474.2959G,2019ApJ...883L..24S, Toffano:2019ekp} indicates that the Universe at high redshift has poor stellar metallicity than at low redshift. As a result, the BHs formed at high redshift may have a higher PISN mass scale than the BHs formed at low.  

(iii) Finally, the BHs which we observe using GWs, are not the individual BHs, but binaries. Though the formation of a black hole takes only a few Myrs, a black hole requires much more time to form a binary and merge. As a result, there is a non-zero delay time between the formation of a star and the merging of BHs. This delay time depends on the formation channels of the BBHs \citep{2010ApJ...716..615O,2010MNRAS.402..371B, 2012ApJ...759...52D, Dominik:2014yma, 2016MNRAS.458.2634M, Lamberts:2016txh, 2018MNRAS.474.4997C, Elbert:2017sbr, Eldridge:2018nop, Vitale:2018yhm,  Buisson:2020hoq,Santoliquido:2020axb}. Moreover, the delay time is not a fixed number for all the BHs, but rather it follows a distribution that is expected to be a power law from simulations. For a flat in the log-space distribution of the separation of the BHs, the delay time distribution is going to be $t_d^{-1}$ with a minimum delay time from a few hundreds of Myrs to a few Gyrs, depending on the formation channels. The current constraints from GWTC-2 \citep{Fishbach:2021mhp} and the stochastic GW background \citep{Mukherjee:2021ags} are weak. In the future, data-driven measurement is possible by combining GW sources with emission line signal \citep{Mukherjee:2021bmw}. Consequently, by combining these three effects, we can expect that the observed BHs detected in a binary system are going to have a redshift-dependent mass distribution due to a phenomena of mixing of BHs. 

\begin{figure}
    \centering
    \includegraphics[scale=0.45]{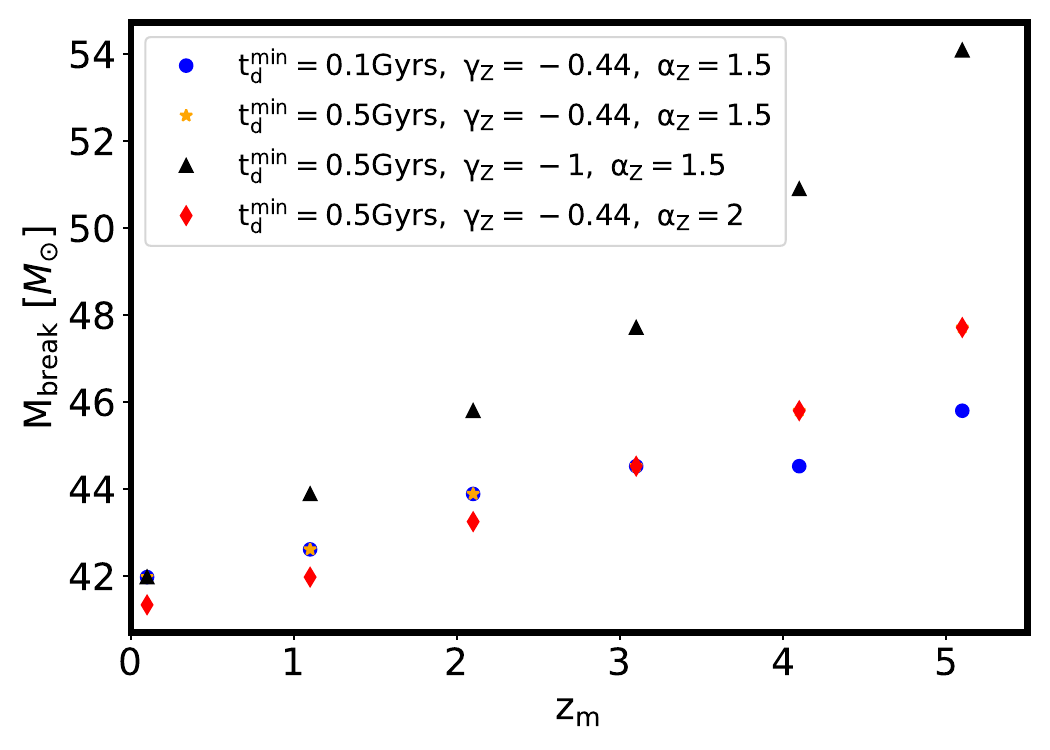}
    \caption{The position of the $M_{\rm break}$ as a function of redshift varying different parameters. The plot was created with fixed $d=-1, \ H_0=70$ km/s/Mpc and $ \ \Omega_m =0.3$. Varying $d$ does not affect the position of $M_{\rm break}$.}
    \label{fig:mpisn}
\end{figure}

\paragraph*{Origin of the redshift dependence of black hole masses:}
Following the analysis that was presented in \cite{Mukherjee:2021rtw} 
 {
the mass distribution of BHs at a merging redshift $z_m$ is given by
\begin{equation}
    p(m_1|z_m,\Phi_{m_1},\Phi_{d},\Phi_{nuis}) = p(m_1|z_m,\Phi_{m_1})W_{t_d}(m_1;z_m|\Phi_{nuis}).
    \label{eqprob}
\end{equation}
where $m_1$ is the most massive BH mass in the source frame, $p(m_1|z_m,\Phi_{m_1})$ is a BH mass distribution which can be associated with the initial stellar mass function (IMF) and $W_{t_d}(m;z_m)$ is the window function that takes into account the delay time of the mergers \citep{Mukherjee_2021} . The BH mass model used in this analysis is described in detail later in Eq. \eqref{eq:pop_model}}.   
The window function given in Eq. \eqref{eqprob} is calculated using
\begin{equation}\label{eq:window_function}
    W_{t_d}(m;z_m) = N \int_{z_m}^{\infty}P_t(t_d|t_d^{min},t_d^{\rm max},d) \frac{dt}{dz_f}W(m;z_f)dz_{f},
\end{equation}
where $N$ is a normalization factor, $P_t(t_d|t_d^{min},t_d^{\rm max},d)$ is the delay time distribution, $W(m;z_f)$ is a Heaviside step function $W(m;z_f)=\Theta(M_{\rm PISN}(z_f) - m)$ and $z_{f}$  is the redshift of the formation of a  black hole which puts a cutoff up to a mass of M$_{\rm PISN}$. The delay time distribution is taken to be a simple power-law function of the delay time $t_d$:
\begin{equation}
    P_t(t_d|t_d^{min},t_d^{\rm max},d) \propto
        \begin{cases}
            (t_d)^{d} & \text{, for $t_d^{min}<t_d<t_d^{\rm max}$}\\
            0 & \text{, otherwise}\\
        \end{cases},
\end{equation}
and the delay time is given by $t_d=t_m-t_f$, with the notation $t_m=t(z_m), t_f=t(z_f)$ to be the time of merger and time of formation respectively. The $W_{t_d}(m;z_m)$ function brings a breaking point $M_{\rm break}$ at the mass distribution, after which the mass distribution is suppressed depending on the form of the delay time distribution, the dependence of the PISN mass scale on stellar metallicity and the redshift evolution of the stellar metallicity. It is evaluated from the combination of different M$_{\rm PISN}$ values which are governed by the minimum delay time $t_{\rm d}^{\rm min}$, metallicity evolution $\gamma_{\rm Z}$ and dependence of PISN mass scale on metallicity $\alpha_{\rm Z}$. The evolution of the $M_{\rm break}$ for different choices of these parameters can be seen in Fig.  \ref{fig:mpisn}.  {At a given $z_m$, the value of $M_{\rm break}(z_m)$ is the minimum of PISN over the formation redshifts included in $z(z_m,t_d^{\rm min})<z_f<z(z_m,t_d^{\rm max})$, namely 
\begin{equation}
    M_{\rm break}(z_m) = {\min}_{t_d \in [t_d^{\rm min},t_d^{\rm max}]} M_{\rm PISN}(z_f(t_d,z_m)).
\end{equation}
}

Assuming the dependence of the PISN mass scale on metallicity $ M_{\rm PISN}(Z)$ studied by \citet{Spera:2017fyx, 2019ApJ...887...53F, Renzo:2020rzx} can be reliably scaled using a parameter $\alpha_{\rm Z}$ \citep{Mukherjee:2021rtw}, we can then model it with metalicity as:
\begin{equation}
    M_{\rm PISN}(Z) = M_{\rm PISN}(Z_*)-\alpha_{\rm Z} \log_{10}(Z/Z_*), 
    \label{eq:mpisn}
\end{equation}
and for a power-law redshift evolution of the stellar metallicity (as supported by the current observations \citep{2010MNRAS.408.2115M, 2012A&A...539A.136S,2012ApJ...753...16K,2013MNRAS.430.2891D,Madau2014}), we can write the redshift dependence of metallicity as $\log{Z}(z) = \gamma_{\rm Z} z + \zeta$. Consequently, the previous equation can be written as:
\begin{equation}
    M_{\rm PISN}(z) = M_{\rm PISN}(Z_*)-\alpha_{\rm Z} [\gamma_{\rm Z} z + \zeta-\log_{10}(Z_*)],
\end{equation}
where  {$\zeta= 10^{-2}$} is taken to be a constant to match the low redshift measurement of the stellar metallicity $Z(z=0) \approx 10^{-2}$ and $Z_* = 10^{-4}$. We select this $Z_*$ value as it is a mid-value inside the range for which PISN dependence has been explored in 1-D stellar mass model \citep{2019ApJ...887...53F}. In our analysis, we sample for the value of $M_{\rm PISN}(Z_*)$. 

The above expression is written in terms of the global evolution of stellar metallicity with redshift. However, at any particular redshift, there is going to be additional variation in the stellar metallicity depending on the property of the host galaxy. So, we would expect variation in the parameter $\gamma_{\rm Z}$ on the property of the host galaxy. However, currently, due to the poor sky location error of the GW sources, we cannot identify the host galaxy and hence cannot model the parameter $\gamma_{\rm Z}$ as a function of the galaxy. As a result, there can be an additional variation in the  $\gamma_{\rm Z}$ that cannot be well modeled currently. Similarly, the parameter $\alpha_{\rm Z}$ controls the dependence of the PISN mass scale on the metallicity \citep{2019ApJ...887...53F}. Hence we treat both these parameters as nuisance parameters in this analysis. We choose wide priors on $\alpha_{\rm Z}, \gamma_{\rm Z}$ which broadly include expected values  {from works cited above.}

\paragraph*{The connection of the observed BBH mass distribution with the PISN mass scale:}
We model the distribution of BBHs in terms of their source  frame masses $m_1,m_2$ (with $m_1 \geq m_2$) and merger redshift $z_m$ of the binary as 
\begin{eqnarray}
p(m_1,m_2|z_m,\Phi_{m},\Phi_{d},\Phi_{nuis})= \nonumber\\ 
p(m_1|z_m,\Phi_{m_1},\Phi_d,\Phi_{nuis})p(m_2|m_1,\Phi_{m_2})S_1 S_2,
\end{eqnarray}
The masses in the detector frame (or redshifted masses) are given by:
\begin{equation}
    m^{det} = (1+z_m)m,
\end{equation}
where $m$ are the masses in the source frame. To capture the mass distribution of BBHs that originate from the BHs mass distribution of Eq. \ref{eqprob},  {we consider $p(m_1|z_m,\Phi_{m_1})$  to be given by a power-law distribution superpositioned with the distribution of a Gaussian peak \citep{Talbot_2018,Abbott2020b,Virgo:2021bbr,LIGOScientific:2021psn}:
\begin{eqnarray}
    p(m_1 |z_m, \Phi_{m_1}) &=& (1 -\lambda_g)P(m_1|M_{\rm min},M_{\rm max},-\alpha)\nonumber \\
    &&+\lambda_{g} G(m_1|M_{break}(z_m),\sigma_g),
    \label{eq:pop_model}
\end{eqnarray} }
where $\Phi_{m_1}=\{M_{\rm min},M_{\rm max},\alpha,\lambda_{g},M_{break}(z_m), \sigma_{g}\}$, $G(m_1|M_{break}(z_m),\sigma_g)$ is a Gaussian distribution with $\mu = M_{break}(z_m)$ and $\sigma=\sigma_g$ and $P(m_1|M_{\rm min},M_{\rm max},-\alpha)$ is a power-law distribution with slope $-\alpha$ between $M_{\rm min}$ and $M_{\rm max}$. In this model the power-law part of the mass distribution is motivated by the power-law form of the initial mass function (IMF) \cite{Kroupa:2002ky} and the Gaussian part of the mass distribution is motivated by the PISN mass scale. The sources merging at redshift $z_m$ due to the contribution from all the higher redshift will lead to an excess near the value of M$_{\rm break}$ and then a decline in the mass distribution due to the window function. 
The position of the Gaussian peak $\mu$ is considered at the break of the window function at that redshift, which depends on the metallicity dependence of the PISN mass scale and delays time distribution. The value of the PISN mass scale is inferred for the metallicity value at $Z_*=10^{-4}$ (for which the results are obtained by \citet{Farmer}).  The Gaussian peak modeled in this analysis gets a physical motivation expected from the PISN mass scale but is also expected to evolve as a function of the redshift of BBHs mergers. 

The distribution of $m_2$ in the source frame is considered to be given by a power law with maximum value $m_1$:
\begin{equation}
    p(m_2 | \Phi_{m_2}) = P(m_2|M_{\rm min},m_{1},\beta).
\end{equation}
Since $m_2$ is conditional to $m_1$, the window function $W_{t_d}$ is being applied also to $m_2$ indirectly. 

Finally, the functions $S_{(1,2)}=S(m_{(1,2)}|\delta_m,M_{\rm min})$ are sigmoid-like window functions to smooth the lower end of the distributions (see appendix A of \citep{Virgo:2021bbr}).
We choose to consider only the position of the Gaussian peak to vary with redshift since this is the most prominent feature in the mass spectrum of BHs and is the best-constrained parameter. Other mass parameters of the mass model ($M_{\rm max}, \alpha, M_{\rm min},...$) can also be given a redshift or metallicity dependence \citep{van_Son_2022}. However currently, with the limited number of GW sources, measurement of the redshift dependence of the additional parameters will be difficult or unlikely to be strongly constraining.

\subsection{The redshift dependence of the BBH merger rate distribution}\label{merger_dist}
The distribution $p(z_m|\Phi_d,\Phi_c)$ takes into account the BBH merger rate as a function of redshift and it is built as
\begin{equation}
p(z_m|\Phi_d,\Phi_c) = C \frac{R(z_m)}{1+z} \frac{dV_c}{dz_m}|_{\Phi_c},    
\end{equation}
where $C$ is a normalization constant, $\frac{dV_c}{dz_m}$ the differential of the comoving volume and $R(z_m)$ the BBH merger rate as function of redshift. The BBH merger rate is built as
\begin{equation}\label{eq:R_z}
    R(z_m) = R_0\frac{\int_{z_m}^{\infty} P_t(t_d|t_d^{min},t_d^{\rm max},d)  R_{SFR}(z_f)\frac{dt}{dz_f}dz_f}{\int_{0}^{\infty} P_t(t_d|t_d^{min},t_d^{\rm max},d)  R_{SFR}(z_f)\frac{dt}{dz_f}dz_f},
\end{equation}
where $z_f$ is the redshift of formation, $R_0$ the BBH merger rate today, $P_t(t_d|t_d^{min},t_d^{\rm max},d)$ is a time-delay distribution between formation and merger of the binary and $R_{SFR}(z_f)$ is a parametrization for the  Madau-Dickinson star formation rate (SFR) \citep{Madau2014}. 

\begin{figure*}
    \centering
    \includegraphics[scale=0.45]{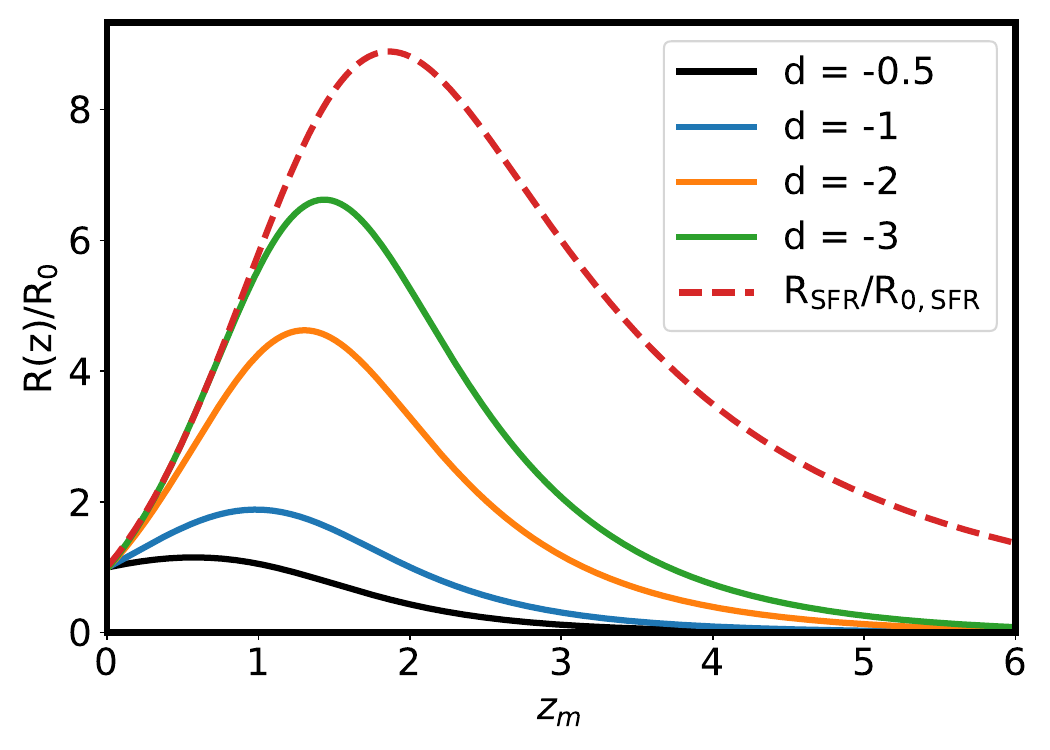}
    \includegraphics[scale=0.45]{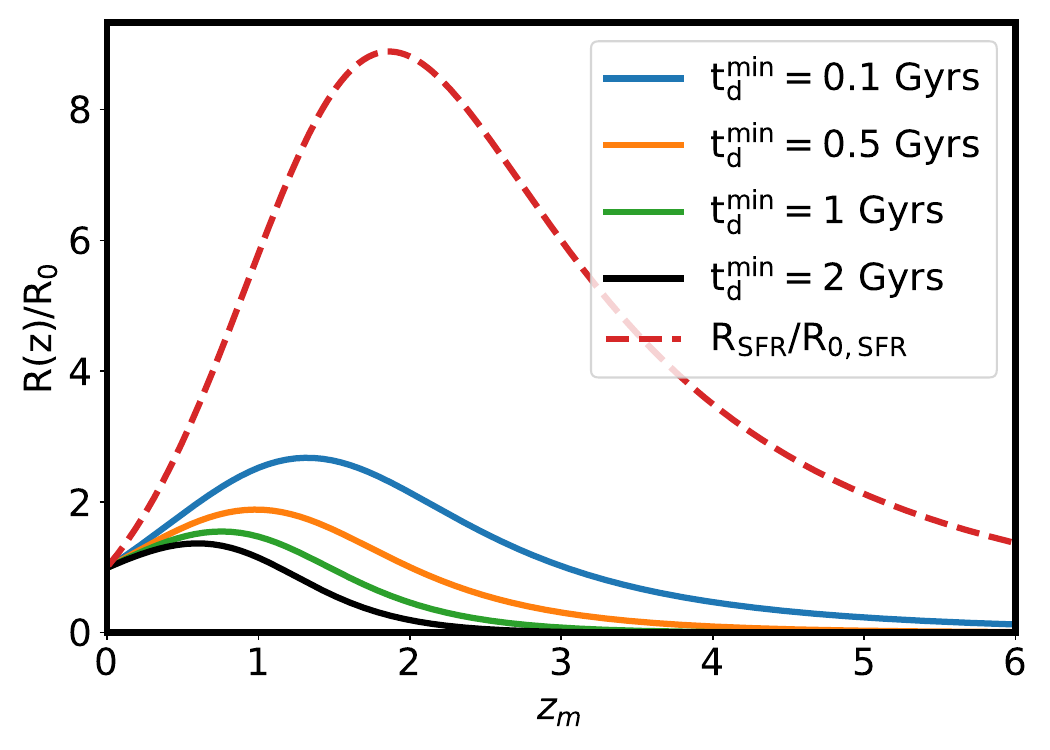}
    \caption{The merger rate function $R(z)/R_0$ for various values of the parameters $d, \  t^{\rm min}_d$ and fiducial flat Cold Dark Matter cosmology with a constant energy density for dark energy, $H_0 = 70$ km/s/Mpc and $\Omega_m = 0.3$. Left: Fixing $t^{\rm min}_d = 0.5$ Gyrs and varying $d$. Right: Fixing $d = -1$ and varying $t^{\rm min}_d$. On the plots, the star formation rate $R_{SFR}/R_{0, SFR}$ can also be seen. }
    \label{fig:R_z}
\end{figure*}

We show the BBH merger rate for a few values of the variables $d, \  t^{\rm min}_d$ and a fiducial flat Lambda Cold Dark Matter (LCDM) cosmology model with $H_0 = 70$km/s/Mpc and $\Omega_m = 0.3$ in Fig. \ref{fig:R_z} with $R_0= 20$ Gpc$^{-3}$ yr$^{-1}$. 
The plot indicates that with the decrease in the value of power-law index $d$ and minimum delay time $t^{\rm min}_d$, the peak in the BBH merger shifts towards a higher redshift with a steeper slope at low redshift. Current observations from LVK can measure BBHs at redshifts ($z<1$) (for a fiducial model of cosmology).

The above discussion shows that the delay time distribution $P_t(t_d|t_d^{min},t_d^{\rm max},d)$ plays a role in both the mass distribution of the BHs and also in their merger rates. As a result, to infer the BBH formation channel and the delay time distribution, it will be necessary to use both merger rate and mass distribution to infer the delay time distribution of BBHs. Moreover, to estimate the redshift of the BHs from their mass distribution, one needs to account for the redshift dependence of the BBH mass distribution. As a result, we need to conjointly infer the cosmological parameters along with the delay time distribution of BHs, and the black hole mass distribution, to correctly marginalize the degenerate parameters.

\section{Bayesian framework to infer cosmology from GW population}\label{bayes}
\begin{figure}
    \centering
    \includegraphics[scale=.35]{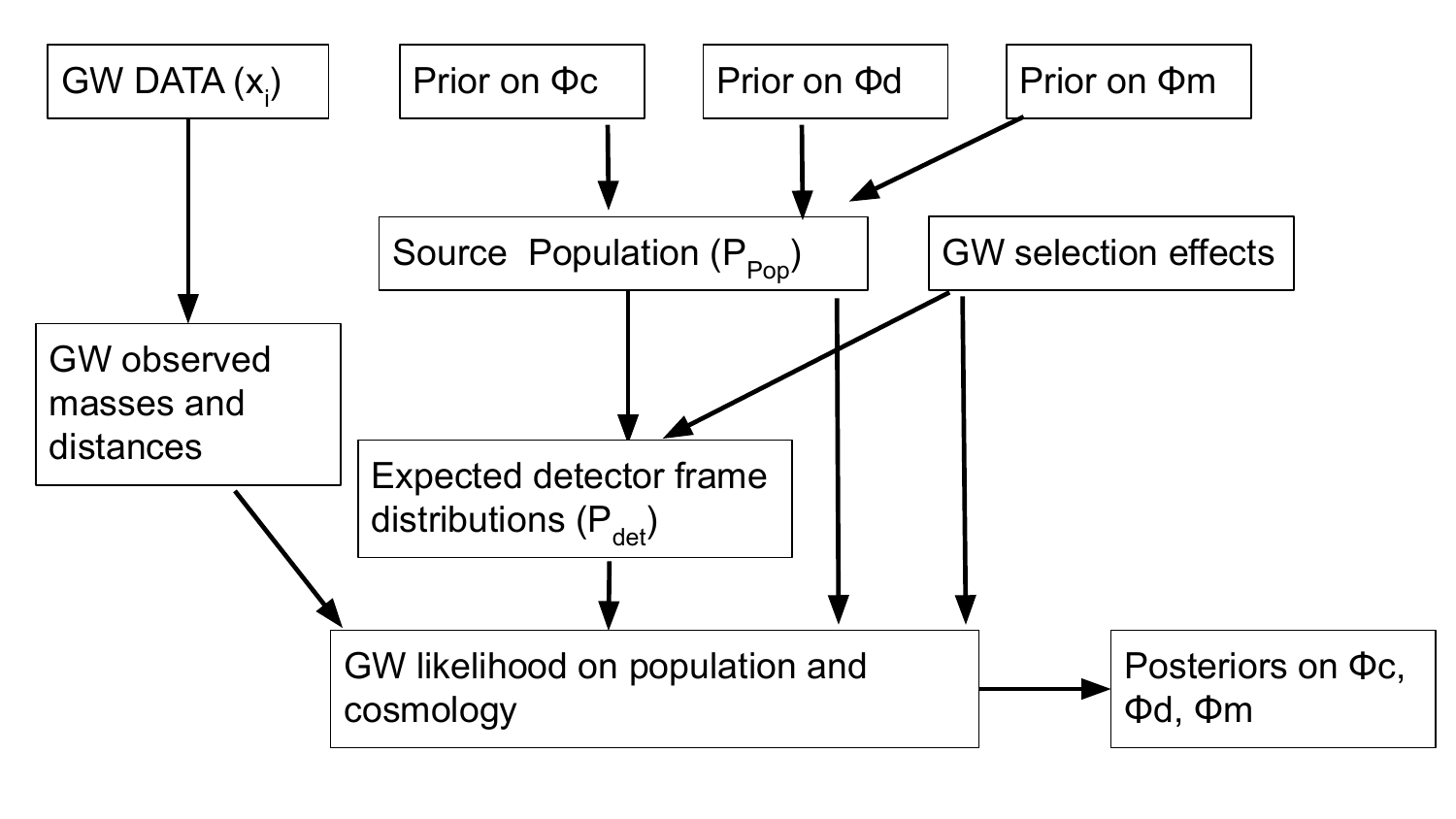}
    \caption{ {Flow chart of the inference method.}}
    \label{fig:flowchart}
\end{figure}
We construct a Bayesian model following the method described in \citep{Mastrogiovanni:2020mvm, Virgo:2021bbr} to conjointly infer the redshift dependence of the mass distribution and merger rate along with the cosmological parameters. In Fig. \ref{fig:flowchart} we show a schematic diagram explaining the formalism. As a cosmological model, we consider a flat LCDM \citep{Aghanim:2018eyx}. Given  a  set  of $N$ GW  detections  associated  with the  data $\{x\}=  (x_1,...,x_{N})$, the posterior on $\Phi$ can be expressed as \citep{Thrane_2019,Mandel_2019,Mastrogiovanni:2020mvm, Virgo:2021bbr,Vitale_2021}
\begin{eqnarray}
\label{eq:posterior}
    p(\Phi, \{x\}, N) &=&\Pi(\Phi) e^{-N_{\rm exp}(\Phi)} N_{\rm exp}(\Phi)^{N_{\rm obs}} \nonumber \times  \\ && \prod_{i=1}^{N}  \dfrac{\int p(x_i | \Phi, \theta) p_{\rm pop}(\theta | \Phi) d\theta} {\int p_{\rm det}(\Phi, \theta) p_{\rm pop}(\theta | \Phi) d\theta}, 
\end{eqnarray}
where $\Pi(\Phi)$ is prior on the parameters, $\theta$ is the set of intrinsic parameters, which are unique for each event,  $p(x_i|\Phi,\theta)$ is the likelihood, $p_{det}(\theta, \Phi)$ is the probability of detection and $p_{pop}(\theta | \Phi)$ is the population modeled prior. Finally, $N _{\rm exp}(\Phi)$ is the expected number of detections in a given observing time, and $N_{\rm obs}$ is the number of events considered in the analysis. The term $p_{pop}(\theta | \Phi)$ is given by Eq. \ref{eq:pop_model_}. This term captures the effects of delay time between formation and merger. 

The denominator of Eq. \ref{eq:posterior} normalizes the numerator and takes into account the selection effects \citep{Mastrogiovanni:2020mvm, Virgo:2021bbr}. It is written as an integral over all detectors' data that pass certain detection criteria for given known noise properties of the detectors. The term $p_{det}(\theta, \Phi)$ is the probability of detecting the source with parameters $\theta$ and assuming hyper-parameters $\Phi$. 

\begin{table*}
  \centering
  \begin{tabular}{cccc}
        {} & \centering{Delay time + Merger rate parameters} & \\
        \hline
        {\bf Parameter} & \textbf{Description} & & \textbf{Prior} \\\hline\hline
        $d$ & Spectral index for the power-law of the delay time distribution. & & $\mathcal{U}$(-4,0) \\
        $t^{\rm min}_d$ & Minimum time for the power-law of the delay time distribution in Gyrs. & & $\mathcal{U}$(0.01,13) \\
        $R_0$ & Value of the merger rate at $z=0$ in $\rm Gpc^{-3} \ \rm yr^{-1}$. & & $\mathcal{U}$(0,1000)\\
        \hline
        \\
        {} & \centering{Mass distribution parameters} & \\
        \hline
        {\bf Parameter} & \textbf{Description} &  & \textbf{Prior} \\\hline\hline
        $\alpha$ & Spectral index for the power-law of the primary mass distribution. &  & $\mathcal{U}$(-4,12)\\
        $\beta$ & Spectral index for the power-law of the mass ratio distribution. &  & $\mathcal{U}$(-4,12)\\
        $M_{\rm min}$ & Minimum mass of the power-law component of the primary mass distribution in $M_\odot$. & &  $\mathcal{U}$(2,10)\\
        $M_{\rm max}$ &  Maximum mass of the power-law component of the primary mass distribution in $M_\odot$.  & &  $\mathcal{U}$(50,200)\\
        $\lambda_{\rm g}$ & Fraction of the model in the Gaussian component. &  & $\mathcal{U}$(0,1) \\
        $M_{\rm PISN}(Z_*)$ & The value of $M_{\rm PISN}$ for the metallicity value $Z_*$ in $M_\odot$.  & &  $\mathcal{U}$(20,60) \\
        $\sigma_{\rm g}$ & Width of the Gaussian component in the primary mass distribution in $M_\odot$.  & &  $\mathcal{U}$(0.4,10)\\
        $\delta_{m}$ & Range of mass tapering at the lower end of the mass distribution in $M_\odot$.  & &  $\mathcal{U}$(0,10)\\
        \hline
        \\
        {} & \centering{Cosmological parameters (Flat LCDM model)} & \\
        \hline
        {\bf Parameter} & \textbf{Description} &  & \textbf{Prior} \\\hline\hline
        $H_0$ & The Hubble constant parameter in km/s/Mpc. &  & $67.4 \rm{(fixed)},\mathcal{U}$(20,150)\\
        $\Omega_m$ & Present-day matter density of the Universe. &  & 0.315 (fixed)\\
        \hline
        \\
        {} & \centering{Nuisance parameters} & \\
        \hline
        {\bf Parameter} & \textbf{Description} &  & \textbf{Prior} \\\hline\hline
        $\alpha_{\rm Z}$ & The parameter that captures a weak logarithmic dependence of $M_{\rm PISN}$ on the metallicity.  & & $\mathcal{U}$(-15,15)\\
        $\gamma_{\rm Z}$ & The parameter that captures the redshift dependence of the metallicity.  & &  $\mathcal{U}$(-5,0)\\
         {$\zeta$} & The parameter that captures the metallicity at redshift $z=0$.  & &  0.01 (fixed)\\
        \hline 
        \hline 
    \end{tabular}
    \caption{
    Summary of the hyperparameters and the priors used. The distribution $\mathcal{U}(min, max)$ is just a uniform distribution between min and max for each parameter.  {Note that the breaking mass $M_{\rm break}(z_m)$ is implied by the choice of the other population parameters.}}
  \label{tab:priors}
\end{table*}
The summary of the priors used for the parameters that we consider in our model can be found in Table \ref{tab:priors}.

\begin{figure*}
    \centering
    \includegraphics[scale=0.25]{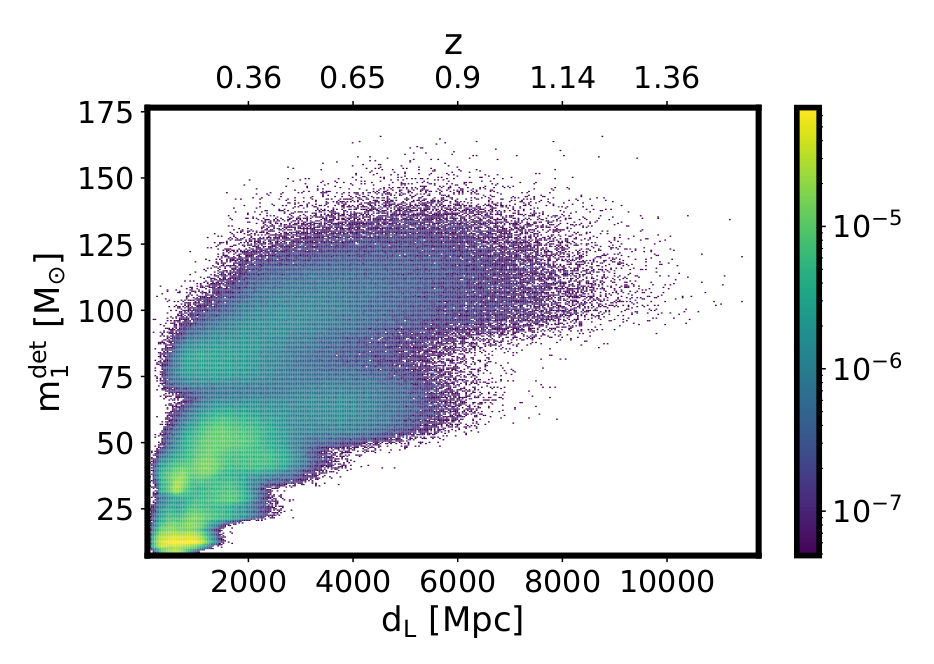}
    \includegraphics[scale=0.25]{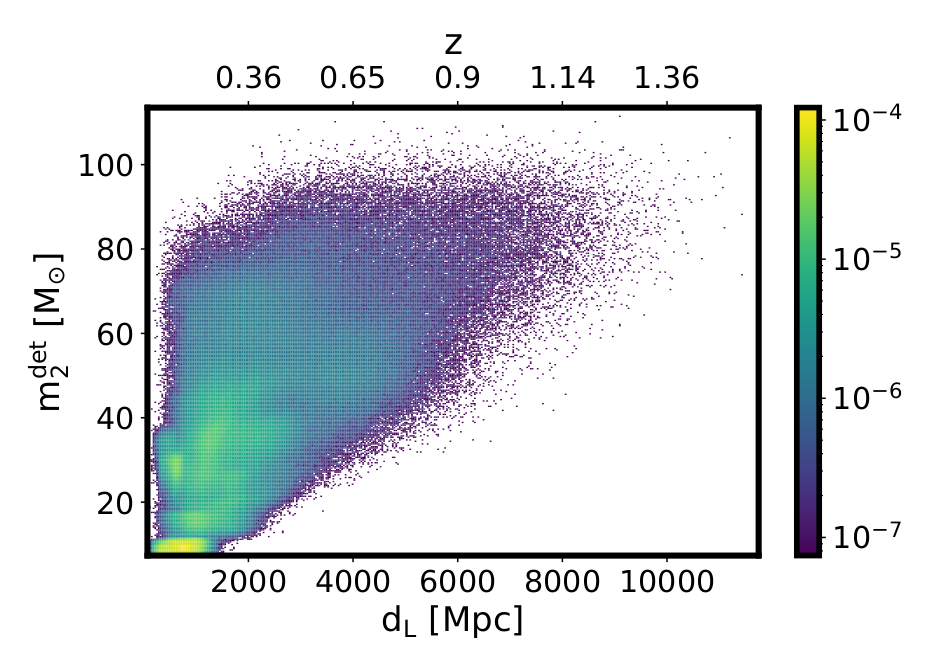}
    \caption{Two-dimensional plots of the detector frame masses and distance posterior samples for all GW events with $SNR\geq 12$.  {In the color bar, the probability in the logarithmic scale can be seen.} Top plot: Samples of the heavy component $m_1^{det}$. Bottom plot: Samples of the light component $m_2^{det}$. }
    \label{fig:mdl}
\end{figure*}
\section{Results from GWTC-3}\label{results}

We analyzed all the BBH events from GWTC-3 \citep{LIGOScientific:2021djp} with the matched filtering Signal-to-Noise Ratio (SNR) in at least one of the detection pipelines higher than 12.  {We select a high SNR threshold to avoid any possible contamination from noise. Moreover, the choice of a high SNR threshold is motivated by the fact that our selection biases are evaluated with injections in simulated and not real data. Detection properties between simulated data and real data might be different, especially when lowering the threshold for detection.} For all the events, we also require a False Alarm Rate $< 0.25 \, {\rm yr^{-1}}$.

The injection campaign is done in simulated  {data with duration and sensitivity typical of O1, O2, and O3}. For all the observing runs, we assume independent duty cycles among the LIGO Hanford (H1), LIGO Livingston (L1), and Virgo (V1) detectors taken from \citep{2021SoftX..1300658A} for O1 and O2 and O3 \citep{Buikema:2020,2023arXiv230203676T}. For O1 we assume duty cycles of 64.6\% for H1 and 57.4\% for L1, while in O2 it was 65.3\%
and 61.8\%. For the entire O3, we assumed 74.6\% for H1, 77.0\% for L1, and 76.0\% for V1. We used the  {power spectral density} of the publicly available detectors for O1\footnote{\url{https://www.gw-openscience.org/O1/}} and O2 \footnote{\url{https://www.gw-openscience.org/O2/}}, while for O3 we used an estimation provided with the first three months of O3a \footnote{\url{https://dcc.ligo.org/LIGO-T2000012/public}}.  {Moreover we have assumed the noise of the detector to be Gaussian and stationary.}
The injections are performed using the  {\textsc{IMRPhenomPv2}} waveform model and are drawn from a distribution in detector frame masses and luminosity distance large enough to cover all the detectable sources assuming a Gaussian stationary noise. The mass and luminosity distance distributions of the GWTC-3 catalog for sources with SNR $\geq 12$ can be seen in Fig. \ref{fig:mdl}. 

For the results in this section, we do not consider GW190521, as it is expected to belong to second-generation BHs (results with the inclusion of this event can be found in \ref{app:3}). A previous study has also found stronger evidence in favor of GW190521 being a second-generation BBH source and other BBH sources did not show up very strong support in favor of the second-generation sources \citep{Kimball:2020qyd}. However, we have shown in this analysis that the constraints on the GW source population parameters including GW190521 are very similar to the case without GW190521 (See \ref{app:3}).

We consider three sets of population priors in this analysis, \textit{Case-1}:  We fix the cosmological parameters besides the Hubble constant and consider only priors on the population parameters describing the BBHs distribution (other cosmological parameters are kept fixed at Planck-2018 cosmology \citep{Aghanim:2018eyx}), \textit{Case-2}: we keep fixed the values of the cosmological parameters to Planck-2018 \citep{Aghanim:2018eyx} and estimate the parameters which are related to the GW source population. We consider this case to infer the value of the GW source parameters assuming a fixed cosmology. Though the choice of cosmological parameters can influence the inferred values of the GW source parameters, given the current precision of the cosmological parameters from Planck-2018, the expected statistical error in the inferred GW source parameters is much larger than the systematic error due to different choices in the value of $H_0= 66.9 $ km/s/Mpc \citep{Aghanim:2018eyx}  or $H_0= 73$  km/s/Mpc  \citep{Riess:2021jrx}. \textit{Case-3}: as Case-2 but keeping the value of the delay time power-law index fixed and equal to $d=-1$, which is usually assumed as a fiducial scenario for flat in the log-space distribution of the separation between the binaries. 

The analysis for Case-3 (with a fixed value of $d=-1$) is motivated to find the constraints on the parameter space for the fiducial scenario of the delay time distribution, which resembles closely previous analysis \citep{LIGOScientific:2021psn}. The priors of the runs for each of the parameters can be found in Table  \ref{tab:priors}. We have summarised the estimated values of the parameters for different cases in Table \ref{tab:estimations}. All the quoted values of the error bars are 68$\%$ confidence intervals unless mentioned otherwise. 

\begin{figure*}
    \centering
    \includegraphics[scale=1.6]{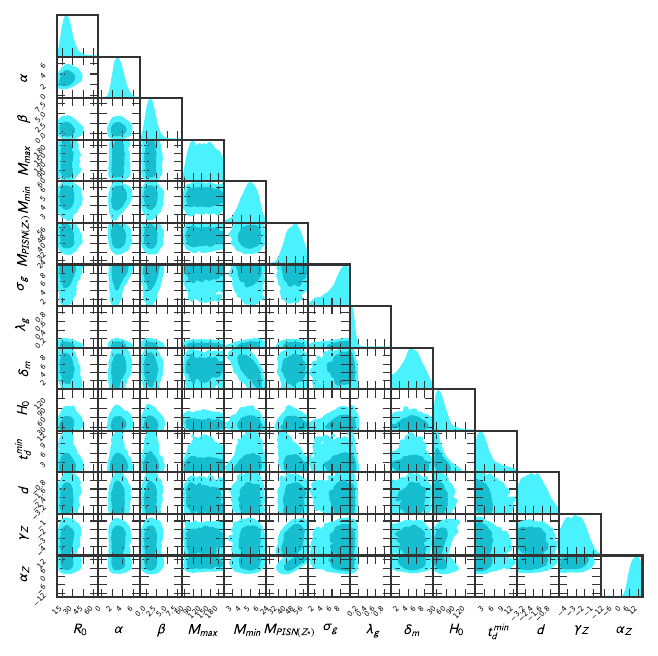}
    \caption{Posterior distributions for all the hyperparameters. Here we have fixed all cosmological parameters besides $H_0$ in Planck-2018 cosmology. We have used all GW events with $SNR\geq 12$. This plot corresponds to case 1 mentioned in the results section.}
    \label{fig:pop+H0}
\end{figure*}

\begin{figure*}
    \centering
    \includegraphics[scale=1.6]{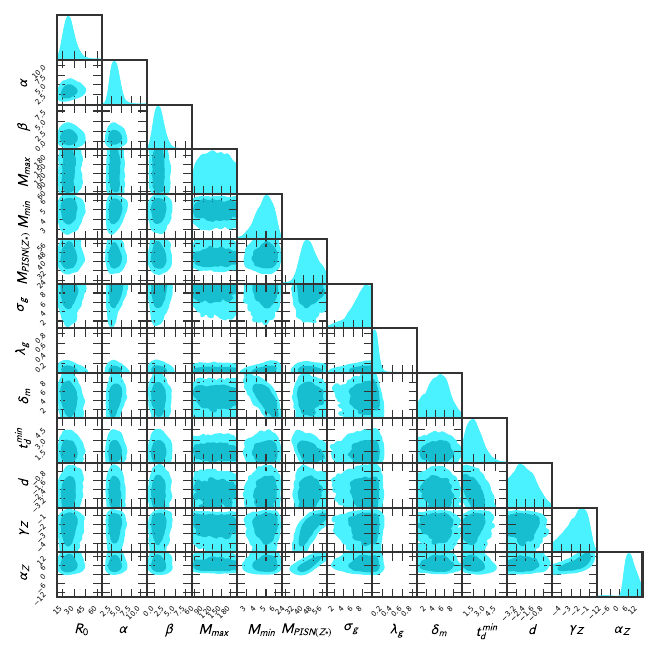}
    \caption{Posterior distributions for all the hyperparameters. Here we have fixed all cosmological parameters at Planck-2018 cosmology. We have used all GW events with $SNR\geq 12$. This plot corresponds to case 2 mentioned in the results section. }
    \label{fig:pop}
\end{figure*}

\begin{figure}
    \centering
    \includegraphics[scale=.5]{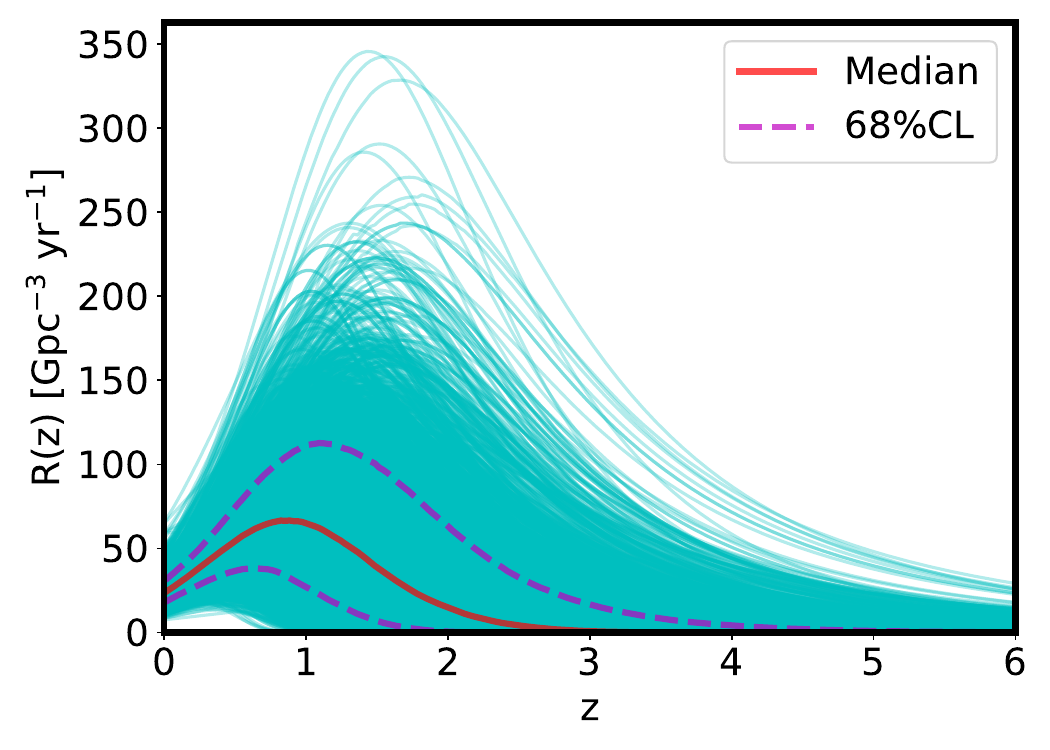}
    \caption{The merger rate evolution as a function of redshift for different posterior samples(cyan curves) of the various parameters. In the same plot, the median (red solid curve) and the $68 \%$ credible levels (purple dashed curves) can also be seen. The cases with a fixed value are shown with the fixed mean value and zero uncertainty.}
    \label{fig:rz}
\end{figure}
\begin{figure}
    \centering
    \includegraphics[scale=.5]{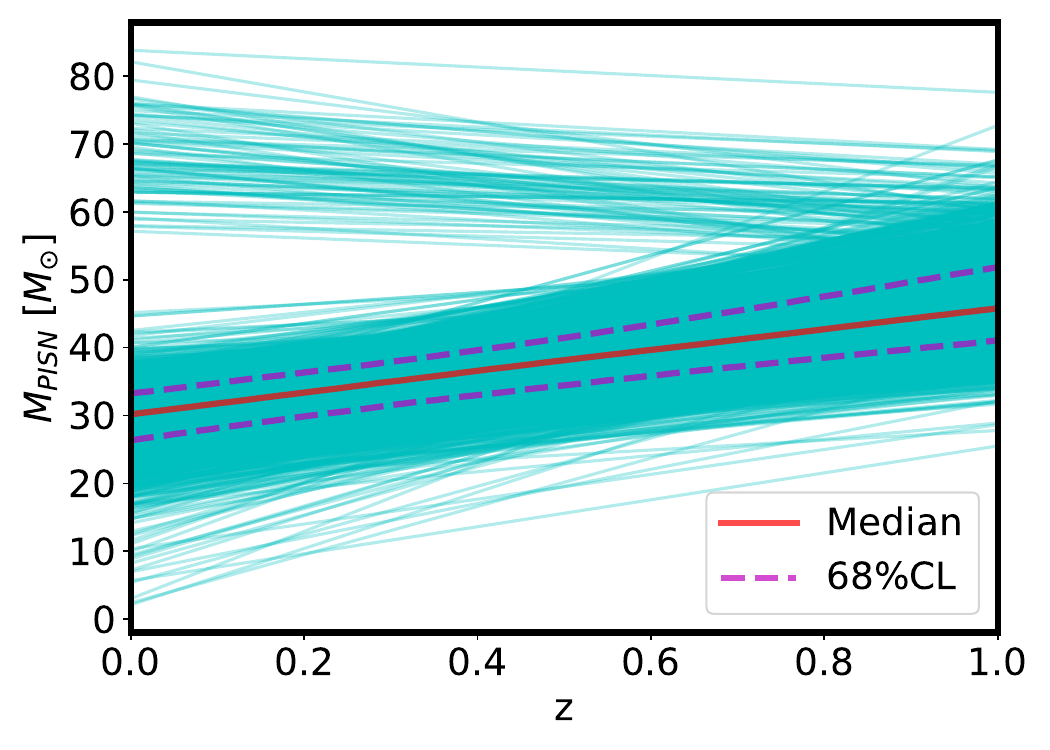}
    \caption{ The $M_{\rm PISN}$ position as a function of redshift for different posterior samples (cyan curves). In the same plot, the median (red solid curve) and the $68 \%$ credible levels (purple dashed curves) can also be seen.}
    \label{fig:mpisn_samples}
\end{figure}

\begin{figure}
    \centering
    \includegraphics[scale=0.45]{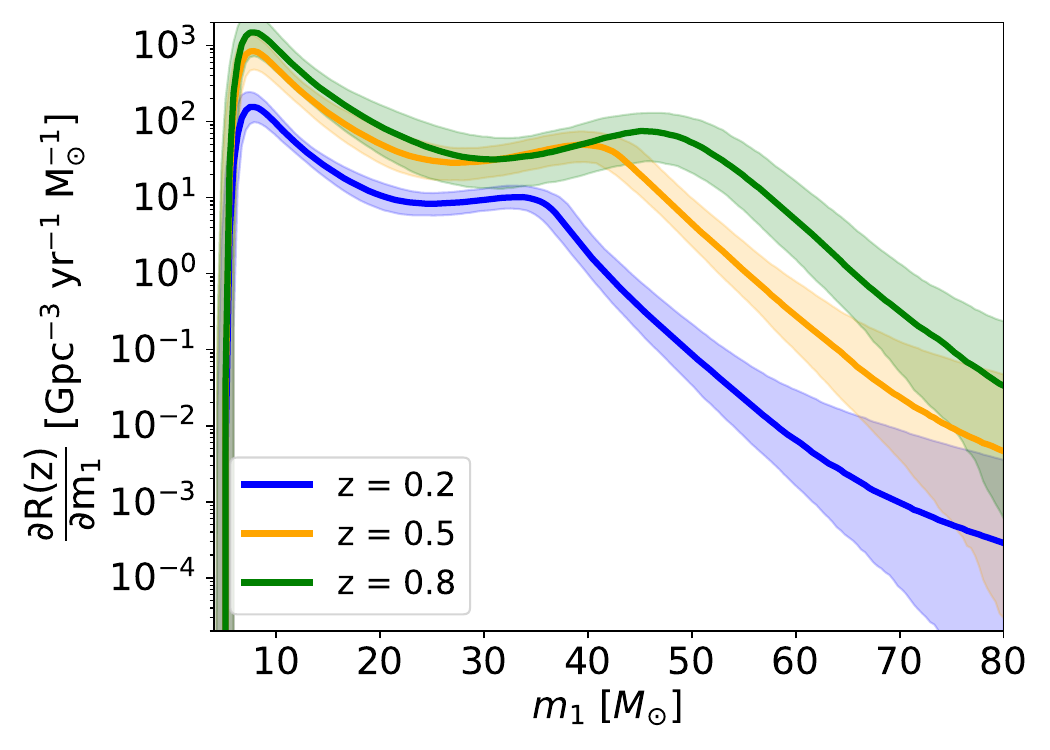}
    \caption{  {Differential merger rate of BBHs over primary mass as a function of redshift}. The different colors indicate the merger rates at different redshifts. Solid lines show the median of the distribution, whereas the shades indicate the 68$\%$ CL.}
    \label{fig:physical_rate}
\end{figure}

\begin{figure*}
    \centering
    \includegraphics[scale=1.6]{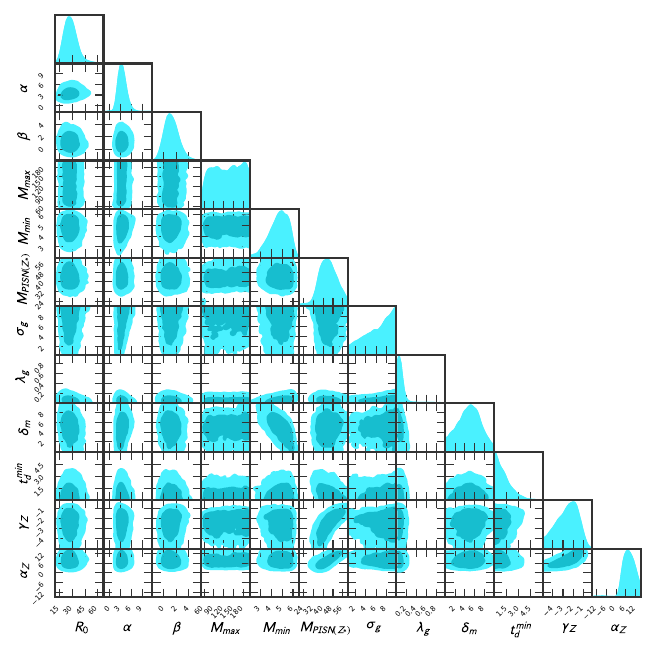}
    \caption{Posterior distributions for all the hyperparameters. Here we have fixed all cosmological parameters at Planck-2018 cosmology. We have fixed the power-law index of the delay time to $d=-1$. We have used all GW events with $SNR\geq 12$. This plot corresponds to case 3 mentioned in the results section. }
    \label{fig:pop+H0+d}
\end{figure*}
\begin{figure}
    \centering
    \includegraphics[scale=.5]{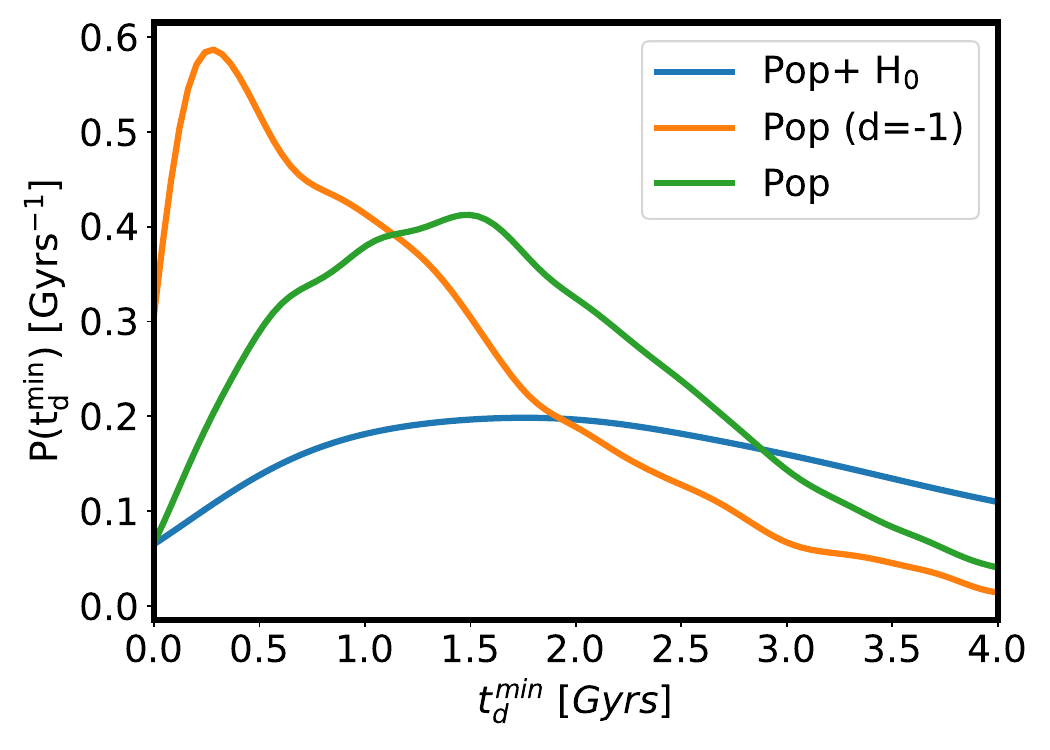}
    \includegraphics[scale=.5]{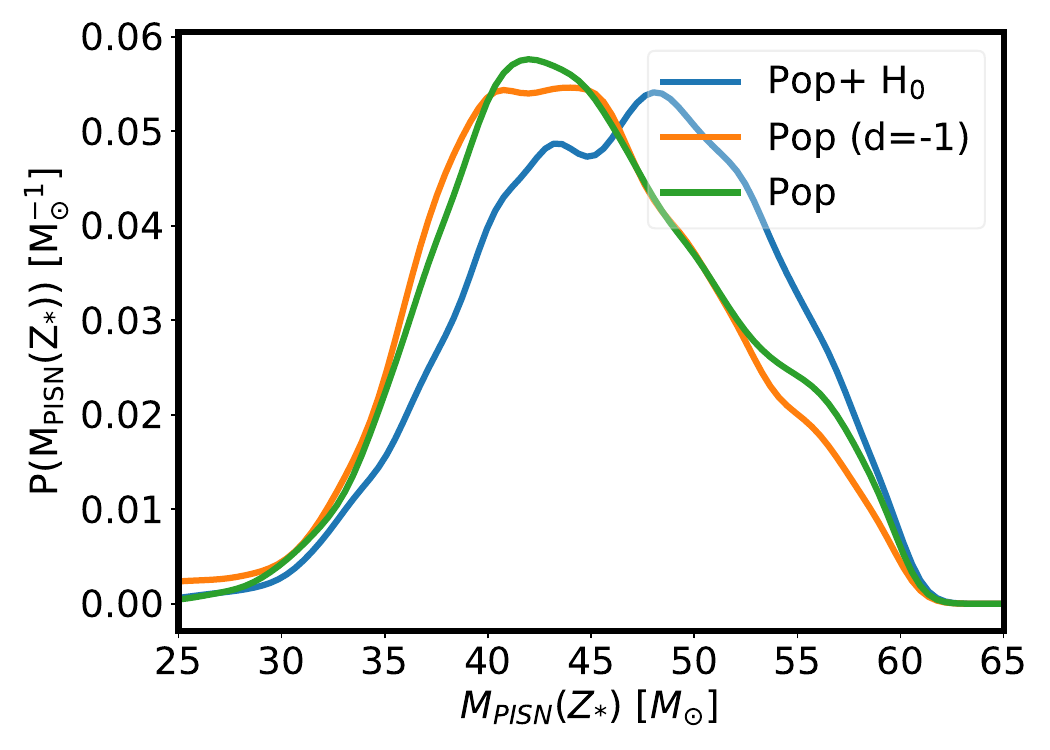}
    \caption{Posteriors on $t^{\rm min}_d$(top) and $M_{\rm{PISN}}(Z_*)$(bottom) for the three cases that we considered. Case 1: Keeping all the cosmological parameters besides $H_0$ fixed to Planck 2018 values and estimating $\rm H_0+$population parameters (labeled as Pop+$\rm H_0$). Case 2:  Fix all cosmological parameters to Planck 2018 values and estimate all of the population parameters(labeled as Pop). Case 3: Keep all the cosmological parameters fixed to Planck 2018 values but also consider a fiducial fixed delay time power-law index $d=-1$ and estimate the rest of the population parameters(labeled as Pop(d=-1)). }
    \label{fig:mugtmin}
\end{figure}

\begin{table*}
\centering
\onecolumn
\input{tablefile}
\caption{The median estimations of all parameters along with the 68$\%$ credible levels can be seen here.  {Note that the table reports values also for parameters that were not constrained in the prior range. For a description of what are the constrained parameters, see the relevant result Sec.~\ref{results}.} The $N_{\rm exp}$ and $N_{\rm events}$ rows indicate the expected number of events in each case using the median value and are derived from the estimated parameters and the number of events detected above an SNR threshold from GWTC-3 respectively. The model that we find to be the most preferred one is highlighted in bold.}\label{tab:estimations}
\twocolumn
\end{table*}

\textbf{Case-1 (GW source population + $H_0$):} For this scenario, the joint constraints on the thirteen GW source population parameters and one cosmological parameter, $H_0$, is shown in Fig. \ref{fig:pop+H0}. The joint estimation can be broadly classified into the parameters related to the delay time + merger rate, mass distribution, and cosmology. Among the delay time + merger rate parameters, we find that data support a scenario of a steep increase in the redshift evolution of the merger rate ($d<-1$), with BBHs merger rate density at $z=0$ {$R_0=22.3^{+7.5}_{-5.7}$} $\rm Gpc^{-3} \ yr^{-1}$. 
Our constraint on the GW merger rate at high redshifts is dominated by our assumptions on the SFR and the constraints we obtain on the time-delay parameters at low redshifts.
However, if there is a different population of BBHs that do not contribute to low redshifts according to the Madau-Dickinson SFR, but contributes to the high redshift such as the Pop-III star, that cannot be constrained from this analysis. The expected number of events after including the detector noise and duty cycle agrees well with the total number of events with SNR $\geq12$ considered in this analysis. 

Using this model, we impose constraints on the minimum delay time distribution $t_d^{\rm min}<10$ Gyrs\footnote{Upper or lower limits are based on the 90$\%$ CL}. 
We find the power-law index of the delay time distribution to be constrained $d<-0.7$ and there is a mild preference towards values lower than $d=-1$ as expected from a simple scenario of flat in the log-space distribution of object separation of the binaries. This measurement shows a steep distribution of the delay time and hints towards scenario formation channels having less probability of a large delay time. 
The constraints on the $t_d^{\rm min}$  and $d$ obtained here are driven by the joint estimation of the merger rate and mass distribution of the BBHs.  {Larger (smaller) values of $t_d^{\rm min}$ or larger (smaller) values of $d$ support higher (lower) delay time values in $P(t_d)$. Mergers of BBH from higher redshifts are supported from the scenarios with large values of $t_d^{\rm min}$ or large values of $d$.} 
The BBHs with heavier component masses in the data are fitted with BHs appearing from a higher redshift with higher PISN masses and a non-zero value of the delay time.

In our analysis, we also constrain models with a  metallicity evolution in the Universe through the parameter $\gamma_Z$. The value of the parameter $-\gamma_{\rm Z}=3.2^{+1.1}_{-1.2}$\footnote{Measurements around the median are based on the 68$\%$ CL} shows that there is likely an evolution of the metallicity of the parent stars. In comparison to the current observations, \citep{2010MNRAS.408.2115M, 2012A&A...539A.136S,2012ApJ...753...16K,2013MNRAS.430.2891D, Madau2014} and also proposed from simulations \citep{2016ApJ...822..107G,2019MNRAS.484.5587T}, the posterior on $\gamma_{\rm Z}$ is consistent, supporting a decrease in the stellar metallicity with redshift. However, depending on the metallicity of the host galaxy,  {the parameter $\gamma_Z$ can have additional dependence on the astrophysical property of the host galaxy \citep{Artale:2019doq, Artale:2019tfl}, and hence can exhibit additional variation from the mean metallicity value of the Universe.} Such effects can show up when more events are available and hence better modeling of the BBHs population will be needed.

The parameters related to the mass distribution are also constrained well using a model. We have obtained a value of the power-law index of the mass distribution $\alpha=3.2^{+1.0}_{-0.8}$ and $\beta=1.0^{+1.2}_{-0.9}$.  {We also find support for a feature over a simple power-law in the mass spectrum of BBH mergers with a relative height of the feature with respect to the power-law component of $\lambda_g<0.13$ and the position of the feature is inherited at a fiducial metallicity from 
 $M_{\rm{PISN}}(Z_*)=46.8^{+6.8}_{-7.3} $ $\rm M_\odot$ at $Z_*=10^{-4}$}. Differently from \citep{LIGOScientific:2021psn}, we are not able to exclude with confidence the value $\lambda_g=0$ (absence of a peak feature).   This is due to the use of selection criteria based on a higher SNR cut instead of an IFAR cut. We verified that with a vanilla PLP model and an IFAR cut of 1 yr as in \citep{LIGOScientific:2021psn}, we can exclude the absence of the peak.)

The position of the peak agrees with the theoretically predicted position of the PISN mass scale between 45-60 $\rm M_\odot$ \citep{2019ApJ...887...53F, Renzo:2020rzx,Baxter:2021swn}.  {In Fig.~\ref{fig:physical_rate}, we show the PISN position is translated to the BBH merger primary mass spectrum when taking into account the full-time-delay model. As we can see from the figure, the PISN mass scale between 45-60 $\rm M_\odot$ is translated to a BBH merger excess at around 35 $\rm M_\odot$ for $z<0.2$. This is compatible with the overdensity of BBHs observed by the LVK in \cite{LIGOScientific:2021psn}. However, as redshift increases this moves to higher masses and appears to become more prominent.}

 {For high-mass BBH, we see a significant increase in merger rate with redshift}, clearly indicating that there is support for a higher merger rate from high masses at higher redshift in comparison to the low redshift. The posterior distribution of the M$_{\rm PISN}$ parameter is shown separately in Fig. \ref{fig:mugtmin} (bottom, blue curve). The PISN mass scale depends on the value of metallicity and this value of $M_{\rm PISN}$ is defined in our analysis at the value of $Z_*= 10^{-4}$, which is in agreement with the parameters chosen in the simulation by \citep{2019ApJ...887...53F}.  {The dependence of the PISN mass scale on metallicity is stronger, i.e. the probable values of $\alpha_Z$ are larger, in this model than is expected from the 1-D stellar evolution models of \cite{2011ApJS..192....3P,2019ApJS..243...10P}}.
 
Also, as it is evident from Fig. \ref{fig:pop+H0} the data has strongly suppressed any negative values of $\alpha_Z$. So, scenarios of a decrease in the PISN mass scale with a decrease in the metallicity are strongly ruled out.

It is important to note that the current theoretical estimation on the dependence of the PISN mass scale is subject to the assumption of the stellar wind models and 1-D stellar evolution code MESA \citep{2011ApJS..192....3P,2019ApJS..243...10P}. To explain the current LVK observation of the GWTC-3 by a first-generation BBH formation scenario, one needs a stronger dependence of the PISN mass scale on the stellar metallicity and higher merger rate of the high mass BHs at high redshift. However, in the future with a higher number of sources, a better understanding of the formation channel of the BBHs will be possible.

Previously redshift dependence of different phenomenological models of BBH mass distribution was explored from GWTC-2 \citep{maya2021}. They found strong evidence for the redshift evolution of the mass model when considering a truncated power law with a sharp cut-off at high masses. However, the data were consistent with both an evolving and a non-evolving mass distribution when they considered a broken power-law model as a mass model. 
Those findings are broadly in agreement with our results.

We find that most of the parameters do not show up significant deviation from previous results  \citep{LIGOScientific:2021psn, Virgo:2021bbr}.
However, differently from \citep{LIGOScientific:2021psn, Virgo:2021bbr}, $\sigma_g$ is not well constrained. The high $\sigma_g$ estimation is an indication that there may not be a Gaussian feature in the mass distribution and the mass distribution can be smeared with an extended distribution in the masses at the higher end.

Finally, a weak measurement of the Hubble constant $H_0=42^{+19}_{-12}$ km/s/Mpc is made which is in the agreement with the values from Planck-2018 \citep{Aghanim:2018eyx}, SH0ES \citep{Riess:2021jrx} due to large uncertainty in the current measurements. However, note the correlation between the Hubble constant and the $t^{\rm min}_d$ parameter in Fig. \ref{fig:pop+H0}.  {This indicates that not being able to correctly infer the PISN mass scale and the value of $t^{\rm min}_d$  can lead to an incorrect inference of the cosmological parameters \citep{Mukherjee:2021rtw}.} 

\textbf{Case-2 (Main model):} In this case, we only focus on the GW source population keeping the value of cosmological parameters fixed at the Planck-2018 \citep{Aghanim:2018eyx}. The corresponding joint estimations of the parameters are shown in Fig. \ref{fig:pop}. From the posteriors, we can obtain the merger rate model for various samples, along with the median. This can be seen in Fig. \ref{fig:rz}. We find that the value of the $M_{\rm{PISN}}(Z_*)=44.4^{+7.9}_{-6.3}$ $\rm M_\odot$, has moved to lower values with respect to the value allowing $H_0$ to vary. The value of $t^{\rm min}_d$ shows a maximum a posteriori around $1.5$ Gyrs and a significantly narrower posterior with respect to \textit{Case-1}. The posteriors for $t^{\rm min}_d$ and $d$ are consistent with the case of varying $H_0$. We also find $d<-0.87$ for this case. We retrieve weak evidence of $\rm M_{PISN}(z)$ evolving with redshift as the $\alpha_Z$ parameter supports positive values. The value of $M_{PISN}(z)$ spans from around 30 $\rm M_\odot$ for $z=0$ up to around 40 $\rm M_\odot$ for $z=1$. As shown in Fig. \ref{fig:mpisn_samples}, the redshift evolution of the PISN mass scale shows a weak variation over the redshift range $z\in [0,1]$. However, more observation will be required to confidently make any detection of the redshift evolution of PISN mass distribution.  {A few samples in Fig. \ref{fig:mpisn_samples} show $\rm M_{PISN}(z)$ values around $80$ M$_\odot$ at redshift $z=0$ and exhibit a decrease in the $\rm M_{PISN}(z)$ with redshift evolution. Those arise from the tail of the posterior distribution due to statistical fluctuations.}

Comparison of the simple PLG peak model with our model we retrieve a Bayes factor ($\rm BF_{\rm PLG/\rm {main\, model}}$) equal to $\rm log_{10}(BF_{\rm PLG/\rm {main\, model}})=0.32$ in favor of the simple PLG peak model.  {So, we conclude that there is a slight but insignificant preference for the PLG model. A higher number of detections is required to obtain
decisive evidence for or against the redshift evolution of the BBH mass distribution.}

We also compare the results with the events from GWTC-3 with SNR $\geq 11$ in Fig. \ref{fig:pop+H0_snrs} (shown in the appendix). The results are consistent with the measurement of the parameters made with events having an SNR $\geq 12$.  In this analysis, we have not considered the GW event GW190521 \citep{LIGOScientific:2020iuh} which has a much higher value of the component masses. Results including GW190521 can be seen in the appendix (see in Fig. \ref{fig:pop+H0_GW190521}). Constraints on the GW source parameters are very similar for both with or without GW190521.

To explore whether the model struggles to fit high masses seen in the data, we also consider a variation of our main model. We refer to this modification as the ``High mass'' model and in this we impose the window function $W_{t_d}$ only in the Gaussian peak of the distribution, leaving the power law intact. For a given redshift, we can fit higher mass events with respect to our main model. However, this model is not physically motivated within the framework of mixing binary black holes scenarios. The posteriors we obtained for the High mass model can be seen in the appendix (see in Fig. \ref{fig:high_mass_model}) and are broadly consistent with the results obtained for our default model.
Moreover, the value of the Bayes Factor in favor of our baseline model with respect to this ``High mass'' model is $\rm log_{10}(BF_{\rm main \, model/\rm {High\, mass}})=0.17$. This implies that both models fit the data equally with a slight preference in favor of the baseline model.  In App.~\ref{app:4} we provide more details about this comparison.

\textbf{Case-3 (GW source population (with fixed $d=-1$) parameters):} In \textit{Case-1} and \textit{Case-2}, we have seen a value of the power-law index significantly away from $d=-1$ (usually considered as a fiducial value \citep{Fishbach:2021mhp, Mukherjee:2021ags}), which is possible for scenarios with flat in log space distribution of the separation between the BHs.  Here we perform a joint estimation with the value of the power-law index fixed at $d=-1$. The corresponding joint estimations of the parameters are shown in Fig. \ref{fig:pop+H0+d}. For this case, the merger rate normalization is  $R_0=27.3^{+7.4}_{-6.5}$ $\rm Gpc^{-3} \ yr^{-1}$, and the constraints on the minimum delay time $t^{\rm min}_d$ have been reduced (though completely in agreement with the value obtained with $d$ varying from the two previous cases). This happens because, for a scenario with a fixed value of the parameter $d=-1$, the peak of the merger rate distribution is shifted towards a lower redshift, as a result by allowing a smaller value of the delay time $t_d^{\rm min}$, the peak of the merger rate position shifts towards a higher redshift.  The position of the Gaussian peak has a very similar value $M_{\rm{PISN}}(Z_*) =43.8^{+7.5}_{-6.5}$ $\rm M_\odot$ with respect to the previous results of \textit{Case-2}.

Among all our time delay models, we find that the preferred model is \textit{Case-2}, namely the case in which cosmology is fixed but the index of the time delay distribution $d$ varies. More interestingly, the \textit{Case-3} model ($d=-1$) is disfavored with respect to \textit{Case-2} by a Bayes factor of $\rm log_{10}(BF_{\rm vary\, d/\rm {fixed\, d}})=0.38$.  {However, this is not enough to claim any statistical evidence and we can not conclude any preference towards against or in favor of $d=-1$.} The posteriors of $t_{d}^{min}$ and $M_{\rm{PISN}}(Z_*)$ can be seen in the top and bottom panels of Fig. \ref{fig:mugtmin} for the three main cases that we considered.

 {The results obtained using our model including the delay time distribution and redshift evolution in the merger rate indicate a value of M$_{\rm PISN}= 44.4_{-6.3}^{+7.9}$}. In this model, the heavy mass BHs are formed from the low metallicity parent stars at a high redshift that has merged at a low redshift due to a delay time distribution function which allows large values of time delay. The Bayes factor in favor of this model in comparison to the phenomenological model of PLG is comparable and cannot be well distinguished at this stage. However, in the future with more data, we will likely be able to distinguish between different scenarios.

One of the major drawbacks of the baseline model considered here is that it only considers first-generation BBHs and not the scenarios where the second-generation BBHs are present. As a result, the presence of heavier masses observed in GWTC-3 is expected to arise from stars with low metallicity. However, in reality, there can be second-generation BBHs that can contribute to the observed population. A successful physics-driven model needs to consider this aspect as well. We will explore this in a future work.  

\section{Conclusion}\label{conc}
The mass, spin, and merger rate of BBHs are a direct probe to infer their formation channels.  {Though the information available from observations of BBH component spins is limited (see
e.g. \cite{LIGOScientific:2021psn}), }we can infer the masses and luminosity distance of several BBHs using the network of LIGO/Virgo detectors. The mass distribution and merger rate of the BBHs are likely to exhibit a redshift dependence due to the dependence on the stellar metallicity and formation channel of BBHs. In this work, we consider a model which considers the redshift dependence of BBH mass distribution due to the mixing of binary black holes and  {the redshift dependence} of merger rate.  {We used a Bayesian analysis and estimated the values of the model's parameters using the latest GW catalog of LVK GWTC-3.}

By fitting this model with the data with both GW source parameters and cosmological parameters. We show that using a time-delay distribution for BBHs and a PISN mass model in redshift, it is possible to obtain a value of $M_{\rm PISN}(Z_*)$ compatible with astrophysical expectations.
This shows that the BBH excess in the mass profile of BHs could likely be reconciled with the PISN mass-scale value of around $40-50$ M$_\odot$ when time delay information is considered. However, if there exist BHs of second-generation that are going to have masses heavier than predicted by the PISN mass scale, then such systems cannot be captured by this model. Though we have found that current data equally favors a model allowing only for first-generation BBH mergers with redshift-dependent mass distribution over the phenomenological redshift-independent PLG mass model. There is also an indication of the redshift dependence of the BBH mass distribution, but this trend is not statistically significant. More observations will be required to better understand the redshift dependence.  {Our analysis also indicates a value of $\alpha_Z= 7.0^{+4.0}_{-2.9}$ for the Case-2 (fixed $H_0$ case). This variation is $3-4$ times larger than the typical value people expect from 1-D stellar simulations and a simplistic prescription of stellar wind models \cite{Farmer}. In the future with more data, if these results hold, then one needs to better understand the dependence of metallicity on $M_{\rm PISN}(Z_*)$.}

In the hypothesis that the BBHs we consider are formed in a stellar binary scenario, we provide an upper limit for the minimum of the time delay distribution.
The delay time distribution agrees with the formation channels explored  {in the literature}. We find support for values of the power-law index of the delay time distribution $d<-1$. The value $d=-1$ is usually considered as the fiducial value for flat in-log space distribution of the spacing between the binaries.  {We retrieve values $d<-1$ for the power index of the time delay distribution. Though we can not exclude with certainty the fiducial scenario $d=-1$, we do find a mild preference for smaller values.}

In our analysis, we also jointly infer the value of the Hubble constant, which currently exhibits a large uncertainty and hence is consistent with the value of the Hubble constant inferred from Planck-2018 and SH0ES. However, one of the important parts is that the Hubble constant shows strong degeneracy with the GW source parameters. In particular, in this paper, we have shown that the estimation of the Hubble constant could also be impacted by the BBHs time delay distribution. As a result, if the inferred value of the GW source population is incorrect, then it can bias the value of the Hubble constant and other cosmological parameter estimation.   

In the future with the availability of more sources from the next observation run, a better measurement of the GW source population along with the cosmological parameters will be possible. This will shed light on the formation channels of the binary systems and the redshift dependence of the BH mass distribution. Also, improvement in the theoretical modeling will be required to capture the underlying distribution of the BBHs from both the first generation and second generations.

\section*{Acknowledgements}
The authors thank Konstantin Leyde for reviewing the manuscript as a part of the LVK review process and for providing useful comments. The authors are also thankful to Thomas Dent for suggesting an interesting case (PL+ G*W) that is added in the appendix of the paper.
C. Karathanasis is partially supported
by the Spanish MINECO under the grants SEV-2016-0588 and PGC2018-101858-B-I00, some of which include ERDF funds from the European Union. IFAE is partially funded by the CERCA program of the Generalitat de Catalunya. S. Mukherjee is supported by the Simons Foundation. Research at Perimeter Institute is supported in part by the Government of Canada through the Department of Innovation, Science and Economic Development and by the Province of Ontario through the Ministry of Colleges and Universities. 
S. Mastrogiovanni is supported by the ANR COSMERGE project, grant ANR-20-CE31-001 of the French Agence Nationale de la Recherche.
This analysis is carried out at the computing facility of the LSC cluster. We acknowledge the use of the following packages in this work: Astropy \citep{2013A&A...558A..33A, 2018AJ....156..123A}, BILBY \citep{Ashton:2018jfp}, Giant-Triangle-Confusogram \citep{Bocquet2016}, IPython \citep{PER-GRA:2007}, Matplotlib \citep{Hunter:2007},  NumPy \citep{2011CSE....13b..22V}, and SciPy \citep{scipy}. The authors are grateful for computational resources provided by the LIGO Laboratory and supported by National Science Foundation Grants PHY-0757058 and PHY-0823459. The authors would like to thank the  LIGO/Virgo/KAGRA scientific collaboration for providing the data. LIGO is funded by the U.S. National Science Foundation. Virgo is funded by the French Centre National de Recherche Scientifique (CNRS), the Italian Istituto Nazionale della Fisica Nucleare (INFN), and the Dutch Nikhef, with contributions by Polish and Hungarian institutes. This material is based upon work supported by NSF's LIGO Laboratory which is a major facility fully funded by the National Science Foundation.

 \section*{Data Availability}
The data underlying this article is available on this website \href{https://www.gw-openscience.org}{https://www.gw-openscience.org}.  

\bibliographystyle{mnras}
\bibliography{paper_draft}

\appendix

\section{Parameter estimation with a different SNR threshold.}\label{app:2}

\begin{figure*}
    \centering
    \includegraphics[scale=1.6]{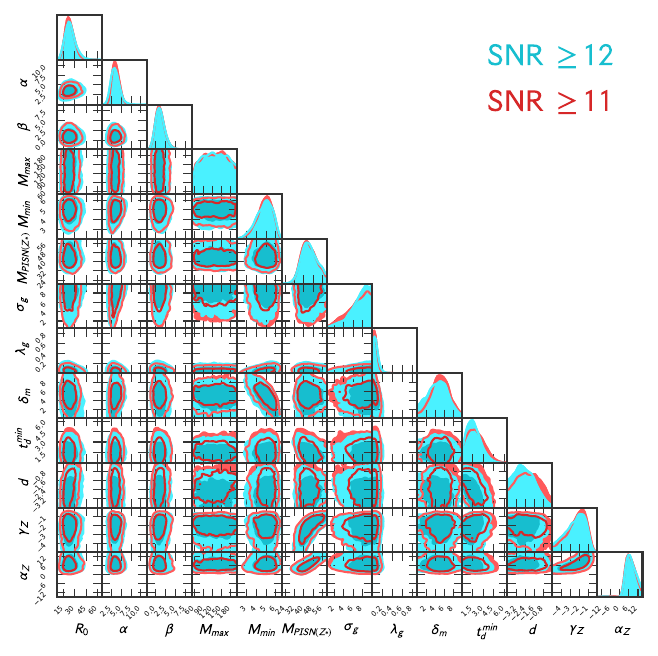}
    \caption{Posterior distributions for all the hyperparameters. Here we have fixed all cosmological parameters at Planck-2018 cosmology. Different posteriors for different SNR thresholds can be seen.}
    \label{fig:pop+H0_snrs}
\end{figure*}

Here we explore the variation of our results in the case of a lower SNR threshold. In addition to the GW events that we used with SNR $\geq 12$ (see Sec. \ref{results}), we are now using additionally all events with SNR $\geq 11$. Lowering the SNR causes to run the analysis with 41 events in total. The different posteriors can be seen in Fig. \ref{fig:pop+H0_snrs}. It is apparent that all of the posteriors are in agreement with the $SNR \geq 12$ results. The only noticeable difference can be seen in the posterior of the time delay's distribution power-law index $d$. We see that the peak of the posterior has moved to slightly higher values, though the posterior is fully in agreement with the SNR $\geq 12$ one. The calculated expected value of the events is $N_{\rm exp}=41^{+6}_{-5}$ and agrees with the 41 events that were used. 

\section{Parameter estimation including GW190521}\label{app:3}

\begin{figure*}
    \centering
    \includegraphics[scale=1.6]{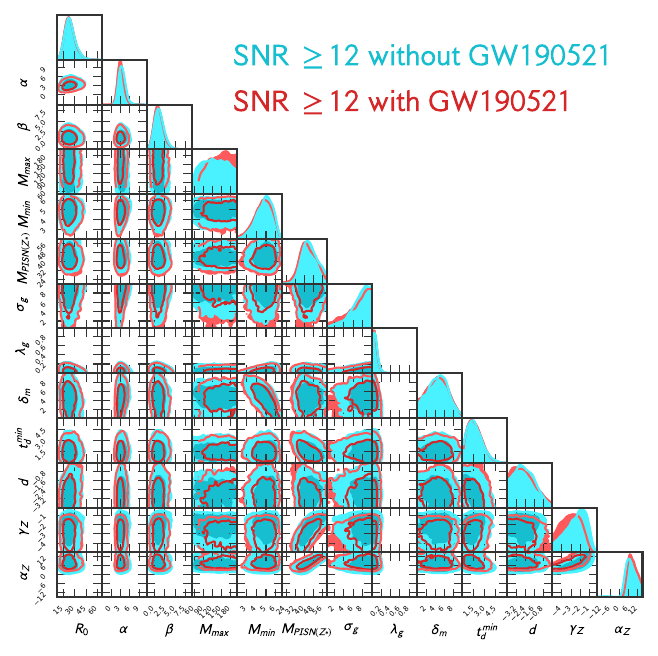}
    \caption{Posterior distributions for all the hyperparameters. Here we have fixed all cosmological parameters at Planck-2018 cosmology. Different posteriors including or excluding GW190521 can be seen.}
    \label{fig:pop+H0_GW190521}
\end{figure*}

The posteriors of all parameters when including the event GW190521 can be seen in Fig. \ref{fig:pop+H0_GW190521}. Again all of the posteriors are in agreement with the ones obtained excluding GW190521. We find a slightly less steep power-law for the $m_1$ distribution with respect to the run excluding GW190521 with a power-law index for the mass distribution of $\alpha=3.0^{+0.9}_{-0.7}$. We also retrieve a flatter $\sigma_g$ posterior and a posterior for $M_{\rm max}$ that disfavors smaller values of masses, though fully in agreement with the previous results. These differences are because GW190521 components are very massive. For the model to fit the extra support at high masses, it needs a less steep power-law index for the mass distribution. At the same time, this leads to a less wide Gaussian peak and to the fact that small values of $M_{\rm max}$ are now disfavoured. We also recover some minor changes in the posteriors for $d, \ \alpha_{Z}$ and $\gamma_{Z}$. We estimate that the expected number of events is $N_{\rm exp}=35^{+6}_{-5}$, which matches the 35 events used.

\section{High mass model}\label{app:4}
\begin{figure*}
    \centering
    \includegraphics[scale=.5]{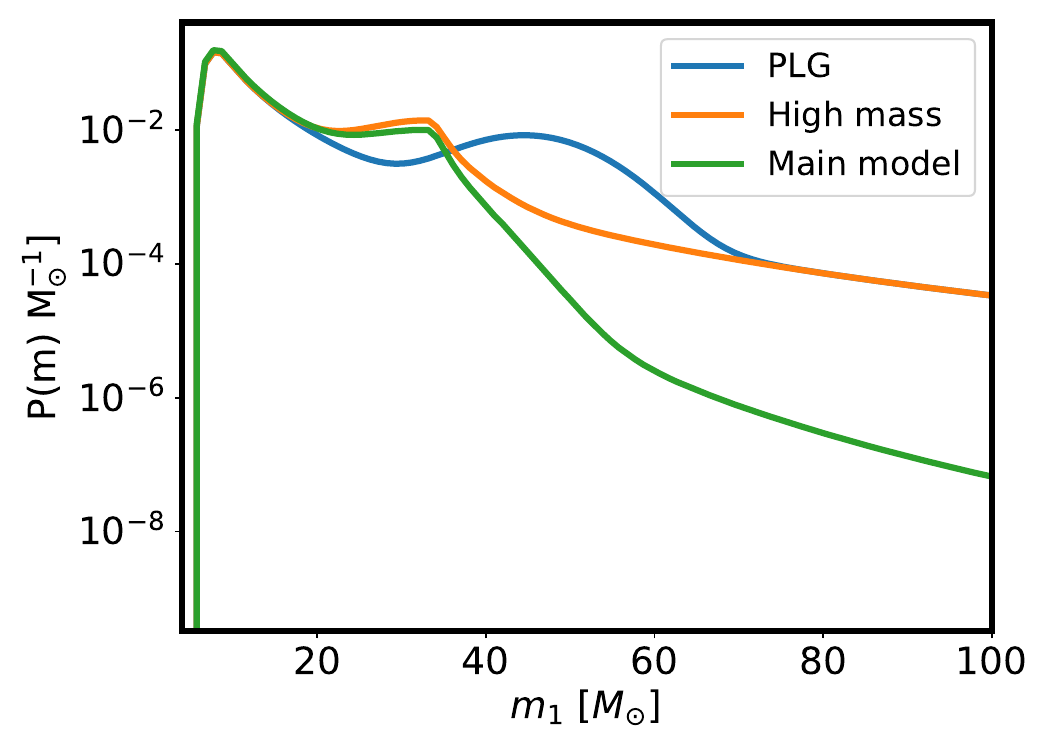}
    \caption{Comparison of the $m_1$ distribution for different mass models. For the values of parameters, we selected the median values of our estimations from \textit{Case-2} at $z_m=0.1$.}
    \label{fig:model_comparison}
\end{figure*}

\begin{figure*}
    \centering
    \includegraphics[scale=1.3]{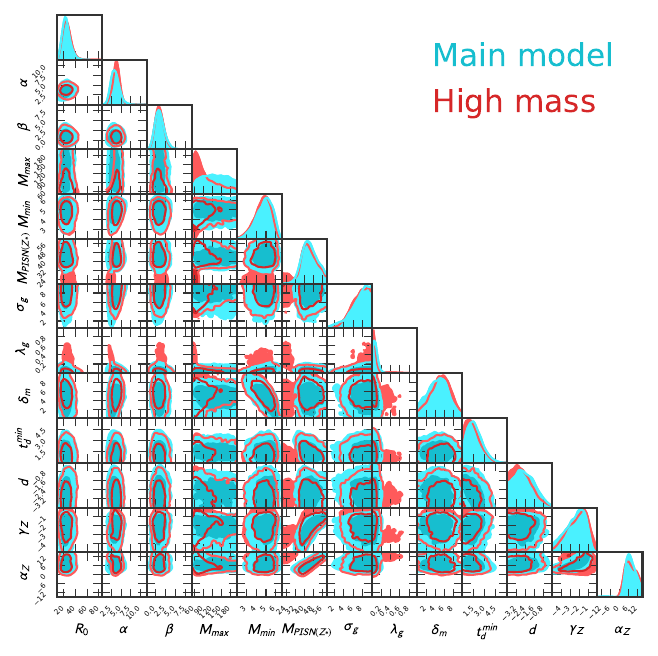}
    \caption{Posterior distributions for all the hyperparameters. Here we have fixed all cosmological parameters at Planck-2018 cosmology. The posteriors for the high mass model can be seen, as well as the posteriors obtained for the usual model.}
    \label{fig:high_mass_model}
\end{figure*}

Here we employ a new model to be able to fit higher masses and we refer to it as the High mass model. Instead of imposing the window function $W_{t_d}$ to the PLG peak, we instead leave the power-law intact and only impose the windowing to the Gaussian peak of the distribution (see Fig. \ref{fig:model_comparison}).  The posteriors obtained by this model can be seen in Fig. \ref{fig:high_mass_model}. It is apparent that most of the posteriors are in agreement with those obtained by the usual model. However, the posterior for $M_{\rm max}$ seems to give more stringent constraints in this case, with high values of masses being disfavoured. This is expected since in this model the power law is left intact allowing for significant support for higher values in the mass distribution after the $M_{\rm PISN}$. Therefore, because we do not have posterior samples in these ranges of masses, the Bayesian inference can exclude the very high mass values. This does not affect our estimations for the time delay parameters, since the posteriors are in agreement with those from our main model.

\label{lastpage}

\end{document}

%% file: tablefile.tex
\begin{longtable}{c|cccccc}%
\hline%
Parameters&Pop+H$_0$&Pop(d=-1)&Pop&Pop inc. GW190521&Pop with SNR${\rm \geq 11}$&High mass\\%
\hline%
${\rm R_0 \ [Gpc^{-3} \ yr^{-1}]}$&$22.3^{+7.5}_{-5.7}$&$27.3^{+7.4}_{-6.9}$&$23.5^{+7.3}_{-5.8}$&$23.2^{+7.2}_{-5.6}$&$23.0^{+6.5}_{-5.0}$&$24.5^{+7.2}_{-6.0}$\\%
${\rm \alpha}$&$3.2^{+1.0}_{-0.8}$&$3.4^{+1.1}_{-0.9}$&$3.4^{+1.1}_{-0.9}$&$3.0^{+0.9}_{-0.7}$&$3.5^{+1.0}_{-0.8}$&$3.7^{+0.8}_{-0.8}$\\%
${\rm \beta}$&$1.0^{+1.2}_{-0.9}$&$1.2^{+1.1}_{-1.0}$&$1.0^{+1.2}_{-1.0}$&$0.9^{+1.2}_{-0.9}$&$1.0^{+1.2}_{-1.0}$&$1.0^{+1.1}_{-1.0}$\\%
${\rm M_{max} \ [M_{\odot}]}$&$129^{+44}_{-43}$&$131^{+48}_{-50}$&$127^{+46}_{-45}$&$144^{+37}_{-42}$&$128^{+45}_{-46}$&$85^{+58}_{-23}$\\%
${\rm M_{min} \ [M_{\odot}]}$&$4.9^{+1.0}_{-0.9}$&$4.8^{+0.8}_{-1.0}$&$4.9^{+0.8}_{-1.0}$&$4.8^{+1.8}_{-1.0}$&$4.9^{+0.7}_{-0.9}$&$4.7^{+0.7}_{-0.9}$\\%
${\rm M_{\rm PISN}(Z_*) \ [M_{\odot}]}$&$46.8^{+6.8}_{-7.3}$&$43.8^{+7.5}_{-6.5}$&$44.4^{+7.9}_{-6.3}$&$42.7^{+8.2}_{-6.5}$&$44.0^{+7.7}_{-6.0}$&$42.8^{+8.4}_{-8.0}$\\%
${\rm \sigma_{g} \ [M_{\odot}]}$&$7.7^{+1.6}_{-3.1}$&$7.0^{+2.2}_{-3.6}$&$7.7^{+1.6}_{-2.6}$&$7.4^{+1.9}_{-3.4}$&$7.2^{+2.0}_{-3.0}$&$7.6^{+1.6}_{-2.1}$\\%
${\rm \lambda_{g}}$&$0.1^{+0.04}_{-0.03}$&$0.1^{+0.1}_{-0.03}$&$0.1^{+0.1}_{-0.03}$&$0.1^{+0.1}_{-0.03}$&$0.05^{+0.04}_{-0.03}$&$0.04^{+0.1}_{-0.02}$\\%
${\rm \delta_{m} \ [M_{\odot}]}$&$4.9^{+2.4}_{-2.6}$&$4.9^{+2.2}_{-2.5}$&$4.7^{+2.2}_{-2.5}$&$4.5^{+2.4}_{-2.6}$&$5.2^{+2.0}_{-2.1}$&$5.3^{+2.1}_{-2.4}$\\%
${\rm H_{0} \ [km \ s^{-1} \ Mpc^{-1}]}$&$42^{+19}_{-12}$&$-$&$-$&$-$&$-$&$-$\\%
${\rm t^{min}_{d} \ [Gyrs]}$&$2.8^{+3.7}_{-1.7}$&$0.9^{+1.2}_{-0.7}$&$1.6^{+1.1}_{-0.9}$&$1.6^{+1.1}_{-0.9}$&$1.9^{+1.4}_{-1.1}$&$1.3^{+1.2}_{-0.8}$\\%
$-d$&$2.4^{+1.0}_{-1.1}$&$-$&$2.5^{+0.9}_{-1.0}$&$2.6^{+0.9}_{-1.2}$&$2.1^{+1.1}_{-1.1}$&$2.8^{+0.8}_{-1.0}$\\%
${\rm -\gamma_{Z}}$&$3.2^{+1.1}_{-1.2}$&$2.5^{+1.3}_{-0.9}$&$2.3^{+1.4}_{-1.0}$&$3.0^{+1.2}_{-1.1}$&$2.2^{+1.3}_{-0.9}$&$2.1^{+1.3}_{-1.0}$\\%
${\rm \alpha_{Z}}$&$9.9^{+3.1}_{-3.4}$&$7.9^{+3.7}_{-3.1}$&$7.0^{+4.0}_{-2.9}$&$8.2^{+3.7}_{-2.8}$&$6.8^{+3.7}_{-2.9}$&$7.1^{+4.5}_{-3.6}$\\%
${\rm{\mathbf{N_{exp}}}}$&$34^{+5}_{-5}$&$34^{+5}_{-5}$&$33^{+5}_{-4}$&$35^{+6}_{-4}$&$41^{+6}_{-5}$&$35^{+5}_{-5}$\\%
${\rm{\mathbf{N_{events}}}}$&$34$&$34$&$34$&$35$&$41$&$34$\\%
\end{longtable}
\addtocounter{table}{-1}